\documentclass[twoside,leqno,twocolumn]{article}

\usepackage[letterpaper]{geometry}

\usepackage{ltexpprt}
\usepackage{hyperref}

\usepackage{times}
\usepackage{soul}

\usepackage{algorithm}
\usepackage{algorithmic}
\usepackage{hyperref}
\usepackage{url}
\usepackage{graphicx}
\usepackage{subfigure} 
\usepackage{amsmath}
\usepackage{amsfonts}
\usepackage{amssymb}
\usepackage{mathrsfs}
\usepackage{booktabs}
\usepackage{multirow}
\usepackage{mathrsfs}
\usepackage{amsmath,amsfonts}
\usepackage[switch]{lineno}
\usepackage[letterpaper]{geometry}
\usepackage{ltexpprt}
\usepackage{chngcntr}
\newcommand{\TheName}{AEDFL}

\begin{document}

\newcommand\relatedversion{}
\renewcommand\relatedversion{\thanks{The full version of the paper can be accessed at \protect\url{https://arxiv.org/abs/1902.09310}}} 

\title{\Large \TheName{}: Efficient Asynchronous Decentralized Federated Learning with Heterogeneous Devices}

\setcounter{footnote}{2}
\author{Ji Liu$^{\dagger*}$\thanks{J. Liu is with Hithink RoyalFlush Information Network Co., Ltd., Hangzhou, China. $^{\dagger}$ equal contribution. $^*$ corresponding author: jiliuwork@gmail.com, yangzhou@auburn.edu.} \and
        Tianshi Che$^{\dagger}$\thanks{T. Che and Y. Zhou are with Auburn University, United States.} \and
        Yang Zhou$^{*\S}$ \and
        Ruoming Jin\thanks{R. Jin is with Kent State University, United States.} \and
        Huaiyu Dai\thanks{H. Dai is with North Carolina State University, United States.} \and
        Dejing Dou\thanks{D. Dou is with Boston Consulting Group, Beijing, China.} \and
        Patrick Valduriez\thanks{P. Valduriez is with Inria, University of Montpellier, CNRS, LIRMM, Montpellier, France and LNCC, Petropolis, Rio de Janeiro, Brazil
.
        }
}


\date{}

\maketitle







\begin{abstract} \small\baselineskip=9pt 
Federated Learning (FL) has achieved significant achievements recently, enabling collaborative model training on distributed data over edge devices. Iterative gradient or model exchanges between devices and the centralized server in the standard FL paradigm suffer from severe efficiency bottlenecks on the server. While enabling collaborative training without a central server, existing decentralized FL approaches either focus on the synchronous mechanism that deteriorates FL convergence or ignore device staleness with an asynchronous mechanism, resulting in inferior FL accuracy. In this paper, we propose an Asynchronous Efficient Decentralized FL framework, i.e., \TheName{}, in heterogeneous environments with three unique contributions. First, we propose an asynchronous FL system model with an efficient model aggregation method for improving the FL convergence. Second, we propose a dynamic staleness-aware model update approach to achieve superior accuracy. Third, we propose an adaptive sparse training method to reduce communication and computation costs without significant accuracy degradation. Extensive experimentation on four public datasets and four models demonstrates the strength of \TheName{} in terms of accuracy (up to 16.3\% higher), efficiency (up to 92.9\% faster), and computation costs (up to 42.3\% lower). 


\end{abstract}

\section{Introduction}

In recent years, a huge amount of data is generated on numerous edge devices, which contain sensitive information of end users, e.g., location information, private images, financial accounts, etc. While the implementation of diverse laws or regulations, e.g., General Data Protection Regulation, hinders the data aggregation, Federated Learning (FL) \cite{mcmahan2021advances} emerges as an efficient approach to deal with distributed data. A typical distributed FL architecture consists of multiple devices and a centralized parameter server \cite{liu2023heterps}, which transfers gradients or models between devices and servers without moving raw data \cite{mcmahan2017communication}. FL is deployed in multiple applications \cite{hsu2020federated,yang2018applied} and various domains \cite{jiang2019networked,li2019privacy}. 

The distributed FL training is typically composed of local training on each device and model aggregation on the server. The server can select available devices and broadcasts a global model to the selected devices. Then, the model is updated based on the local data within each selected device, which is coined local training. After receiving the updated models from the selected devices, the server aggregates them with the global model and generates a new global model, which is denoted as model aggregation. The training process can be either synchronous \cite{mcmahan2017communication} or asynchronous \cite{Su2022How}. With the synchronous FL mechanism, the model aggregation is carried out after receiving all updated models, while, the asynchronous FL enables model aggregation when parts of the models are received. 

While edge devices are generally heterogeneous with diverse computation or communication capacities, the synchronous distributed FL corresponds to inferior efficiency due to stragglers, i.e., modest devices \cite{bonawitz2019towards,park2021sageflow}. The asynchronous mechanism in FL may lead to inferior accuracy or even fail to converge with non-Independent and Identically Distributed (non-IID) data \cite{Su2022How}. In addition, the centralized FL incurs severe communication or computation workload on the server, which becomes a bottleneck and results in low efficiency and a single point of failure \cite{vanhaesebrouck2017decentralized}.

Decentralized FL \cite{qu2021decentralized} is proposed to alleviate the communication bottleneck on the central server. Decentralized FL organizes the devices with a connected topology and enables each device to communicate with its neighbors in a peer-to-peer manner. Each device aggregates its local model with the models or gradients transferred from its one-hop neighbors without relying on a central server. Decentralized FL generally inherits from decentralized learning \cite{ye2022decentralized}, which can be either synchronous \cite{lian2017can,ying2021exponential} or asynchronous \cite{lian2018asynchronous,wu2017decentralized}. The synchronous mechanism relies on a global clock to synchronize the training process on each device, which corresponds to low efficiency with heterogeneous devices. The asynchronous mechanism enables model aggregation without synchronization, which can well utilize heterogeneous devices while incurring a staleness problem and degrading the accuracy. Existing decentralized approaches generally exploit static weights within the model aggregation process with inferior accuracy.

As the computation and communication capacities of edge devices are limited, model compression methods, e.g., pruning \cite{Zhang2022FedDUAP} or sparse training \cite{bibikar2022federated} methods, can be exploited to shrink the models so as to reduce the computation and communication costs. Personalized sparse training in decentralized FL \cite{Dai2022DisPFLTC} is proposed to reduce communication and computation costs with heterogeneous devices. However, the pruning process either degrades the accuracy due to lossy strategies or requires a centralized server. In addition, the decentralized personalized sparse training approach only focuses on the current weights and the local data without considering the interaction with neighbors, which corresponds to inferior accuracy. 

In this paper, we propose a novel \underline{A}synchronous \underline{E}fficient \underline{D}ecentralized \underline{F}ederated \underline{L}earning framework, i.e., \TheName{}, for heterogeneous environments. To deal with heterogeneous devices, we enable asynchronous training on each device with a new dynamic model aggregation method. The dynamic model aggregation method consists of a reinforcement learning-based model selection method and a dynamic weight update strategy. In addition, we propose an original adaptive sparse training method to further reduce computation and communication costs so as to improve efficiency with a lossless method based on the consideration of the impacts on both the current loss and the whole training process. The main contributions are summarized as follows:
\begin{itemize}
    \item We propose an original asynchronous decentralized FL system model with a novel dynamic model aggregation method for collaborative model training with heterogeneous devices. Our proposed dynamic model aggregation method consists of a reinforcement learning-based model selection approach to choose proper models and a dynamic weight update strategy to adjust the weights of each model, which can improve the accuracy. 
    \item We propose a new adaptive sparse training method to reduce the computation and communication costs so as to improve efficiency. The adaptive sparse training method shrinks the model based on the weights of neurons in the model, the gradients, and the values in the Hessian 
    matrix while minimizing the impact on the current loss function to improve the accuracy and considering the impact of the pruning operation in the following training process.
    \item We conduct extensive experiments to compare \TheName{} with representative approaches based on four typical models over four real-world datasets. Experimental results demonstrate the superb advantages of \TheName{} in terms of accuracy, efficiency, and computation costs.
\end{itemize}

The rest of the paper is organized as follows. In Section \ref{sec:relatedWork}, we present the related work. In Section \ref{sec:sysModel}, we present the system model of \TheName{}. In Section \ref{sec:modelAggre}, we propose our dynamic model aggregation method. In Section \ref{sec:sparseTraining}, we propose our sparse training method. In Section \ref{sec:exp}, we demonstrate the experimental results. Section \ref{sec:con} concludes.

\section{Related Work}
\label{sec:relatedWork}

FL \cite{mcmahan2017communication} is proposed to train a global model with distributed non-IID data on  heterogeneous devices \cite{bonawitz2019towards}. Numerous model aggregation algorithms \cite{mcmahan2017communication,Li2020FedProx,Wang2020Tackling,Karimireddy2020SCAFFOLD,Zhang2022FedDUAP,liu2022distributed,liu2022multi,che2022federated,che2023fast,liu2023distributed,che2023federated,JinAccelerated2022,Li2022FedHiSyn,liu2022Efficient} are proposed for synchronous FL. The synchronous FL mechanism is inefficient due to stragglers 
devices \cite{bonawitz2019towards}. 
While asynchronous FL \cite{xie2019asynchronous,huba2022papaya,Liu2024FedASMU} can deal with the device heterogeneity, the staleness may degrade the efficiency or the accuracy \cite{Su2022How}. Although staleness-based weight discount \cite{wu2020safa}, feature representation adjustment \cite{Chen2020Asynchronous}, and learning rate adjustment \cite{WeiStaleness} can adjust the training process, they do not consider the direct impact on the loss function or the difference among diverse devices, which results in inferior accuracy. 

Decentralized FL \cite{liu2022decentralized,qu2021decentralized,hegedHus2019decentralized} enables devices to communicate with their 
one-hop neighbors in a peer-to-peer manner without a central server. Many decentralized FL \cite{ye2022decentralized} approaches directly exploit the decentralized learning techniques \cite{lian2017can,yuan2016convergence}. Synchronous \cite{lian2017can} decentralized learning synchronizes the training process on each device, which favors homogeneous environments. Asynchronous decentralized learning \cite{lian2018asynchronous,wu2017decentralized} can deal with heterogeneous resources while incurring staleness problems. Existing decentralized FL approaches \cite{ye2022decentralized} are generally synchronous while exploiting static weights within the model aggregation process with inefficiency or accuracy degradation. Asynchronous decentralized FL \cite{liu2022asynchronous,li2021blockchain} can deal with heterogeneous devices, while each device needs to send its models to all other devices. This mechanism generates heavy communication overhead when the number of devices is significant. To reduce the communication overhead, exponential topology \cite{assran2019stochastic}, where each device is connected to the magnitude of log($n$) with $n$ representing the number of devices, can be exploited thanks to its
excellent performance \cite{ying2021exponential}. 
Although grid and ring topologies can be exploited as well, they correspond to low generalization capacity \cite{zhu2022topology}.


Pruning techniques \cite{huang2022achieving} can be exploited for sparse training in the FL \cite{babakniya2022federated,bibikar2022federated}, aiming to reduce computation and communication costs for devices with limited computation and communication capacities. 
However, existing sparse training approaches \cite{Dai2022DisPFLTC,huang2022achieving}, which only focus on the weights or ranks of neurons in the model, often lead to reduced accuracy. In addition, the pruning process is usually implemented on models that are not fully trained\cite{babakniya2022federated}, yet existing approaches, e.g., HAP \cite{yu2022hessian}, assume pruning a well-trained model. Furthermore, the pruning process can lead to personalized models \cite{vahidian2021personalized,huang2022achieving,Dai2022DisPFLTC}. However, current methods often overlook the potential impact of the pruned sections on subsequent training, leading to reduced accuracy. Finally, contemporary sparse training methods, e.g., HAP and DisPFL \cite{Dai2022DisPFLTC}, neglect the impact of the pruning process on the gradients, resulting in significant accuracy degradation during FL's training process.

\begin{figure}[!t]
\centering
\includegraphics[width=\linewidth]{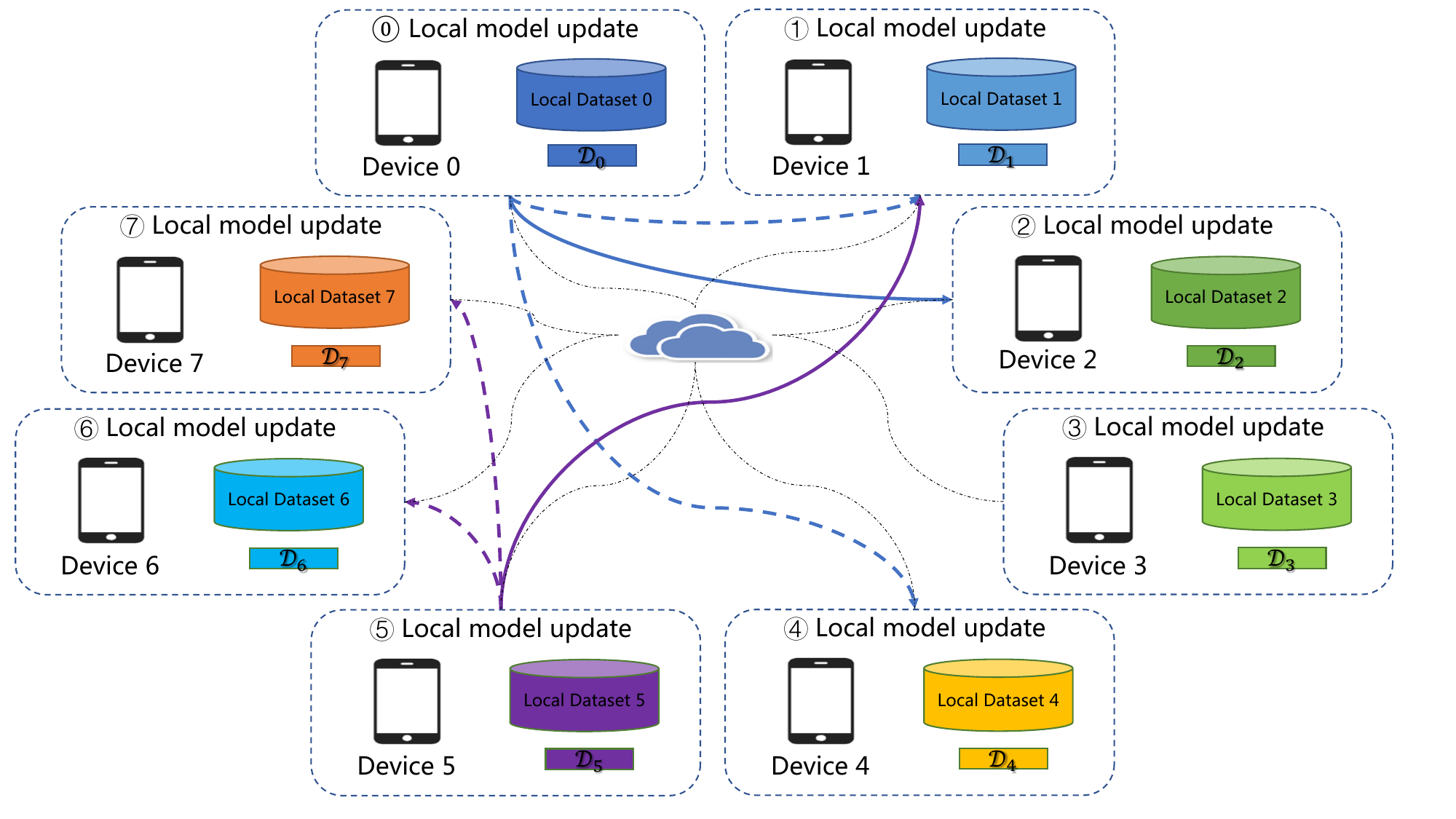}
\vspace{-10mm}
\caption{The system model of \TheName{}.  We consider 8 devices in this figure. Each device has 3 neighbor devices (represents by dashed lines). When a device updates its local model, it sends the model to a randomly sampled neighbor (represented by the solid lines). Each device is coordinated by a coordinator (represented by the dash-dotted lines). }
\vspace{-7mm}
\label{fig:systemModel}
\end{figure}

\vspace{-4mm}
\section{System Model}
\label{sec:sysModel}

In this section, we propose the system of \TheName{}. First, we detail the system architecture of \TheName{}, including the asynchronous communication based on the exponential topology. Then, we present the local update within each device. 

As shown in Figure \ref{fig:systemModel}, we consider a decentralized FL environment with multiple devices and a coordinator. Please note that although there is a centralized coordinator in the system, it is quite different from a central server. Similar to that in Cassandra \cite{lakshman2010cassandra}, the coordinator only manages the index and the heartbeats of each device without participating in the training process or the model aggregation process of FL. Each device is connected to the system and gets an index from the coordinator for the following training process. Each device $i$ has a local dataset $D_i = \{x_i, y_i \}^{s_i}$ with $x_i$ and $y_i$ representing a sample and $s_i$ representing the number of samples on Device $i$. We denote the number of all the samples by $s$. Then, the objective of the training process of FL is formulated as follows:

\vspace{-6mm}
\begin{equation}
\min_{m}\left[\mathcal{F}(m)\triangleq\frac{1}{s}\sum_{i = 1}^n s_i F_i(m_i)\right],
\vspace{-3mm}
\label{eq:problem}
\end{equation}
where $m$ is the parameters of the whole global model (without pruning), $F_i(m_i)\triangleq\frac{1}{s_i}\sum_{\{x_{i},y_{i}\} \in \mathcal{D}_i} f(m_i,x_i,y_i)$ is the loss function on Device $k$ with $f(m_i,x_i,y_i)$ capturing the error of the local model $m_i$ on the sample $\{x_{k},y_{k}\}$. 

\begin{figure}[!t]
\centering
\includegraphics[width=\linewidth]{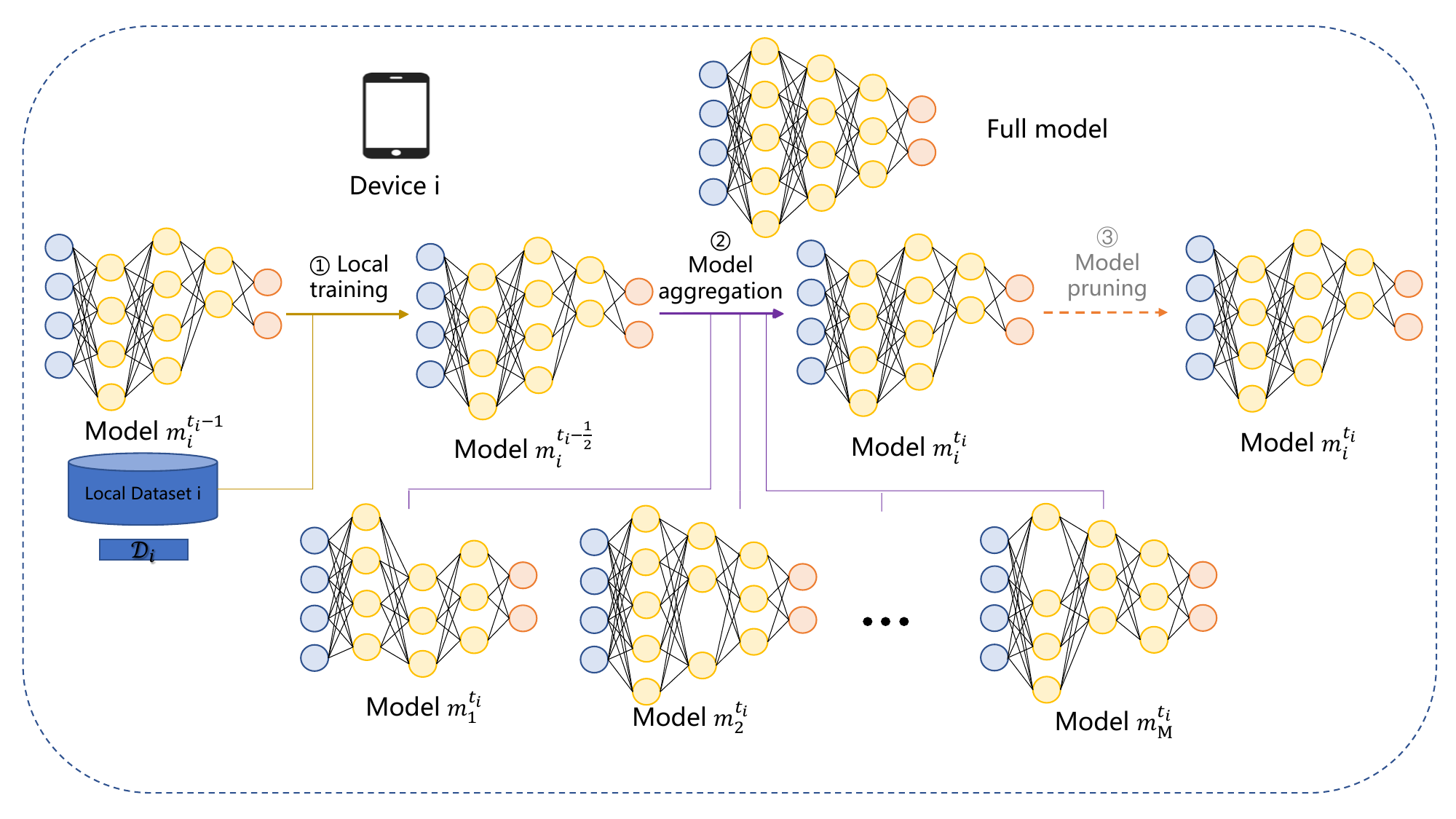}
\vspace{-10mm}
\caption{The local update of \TheName{}. In each device, the models of all its neighbors are cached. When a neighbor sends its updated model, the corresponding cached model is atomically updated. The model pruning (marked in gray) is carried out in specific local updates. The full model structure is kept during the training process. ``M'' represents the number of neighbor models. }
\vspace{-7mm}
\label{fig:localUpdate}
\end{figure}

During the training process, we exploit an exponential topology \cite{assran2019stochastic,lian2017can}, where each device has $\mathcal {O}(\log (n))$ neighbors with $n$ representing the number of devices in the system. Please note that the topology is independent of communication network among devices. The topology represents the information transfer among multiple devices. We assume that the devices can communicate with each other. While the exponential topology is exploited in our framework, other topologies can be utilized as well. We define $w_{i,j}$ as the weight to transfer information from Node $j$ to Node $i$, as follows:
\vspace{-3mm}
\begin{align}\label{wij}
w_{i,j}
\begin{cases}
> 0 & \mbox{$i=j + 2^k$, $k \in \mathbb{Z}$, or $i=j$;} \\
= 0 & \mbox{otherwise.}
\end{cases}
\vspace{-3mm}
\end{align}
In addition, we further define the topology matrix $\mathcal{W} = [w_{i,j}]_{i,j=0}^{n-1} \in \mathbb{R}^{n\times n}$ as the matrix to represent the topology. The training process of \TheName{} consists of three stages. First, each device gets its index from the coordinator. Then, each device performs its local update. Afterward, each device sends its updated model to a randomly selected neighbor. The second and third steps are repeated until satisfying predefined conditions, e.g., a predefined number of iterations on each device is achieved or the consensus distance is smaller than a predefined value. The consensus distance is defined as the average discrepancy of local models between any two nodes. Although each device has $\mathcal {O}(\log (n))$ neighbors, we take advantage of random selection for the model diffusion so as to further reduce communication overhead. Please note that the local update and the model diffusion can be executed in parallel in order to improve efficiency. Furthermore, the local update and the model diffusion are asynchronous and thus independent. The asynchronous training can well alleviate the inefficiency brought by the heterogeneity of devices.


As shown in Figure \ref{fig:localUpdate}, the local update of \TheName{} on each device consists of three steps. First, the local model is updated based on the local dataset exploiting Stochastic Gradient Descent (SGD) \cite{zinkevich2010parallelized}. Afterward, the local model is aggregated with the cached models from its neighbors in the model aggregation process (see details in Section \ref{sec:modelAggre}). To reduce the model size, in certain local updates, we perform model pruning (see details in Section \ref{sec:sparseTraining}). Finally, an updated model is generated, which is sent to one of its neighbors. At the same time, we create another copy of the updated model, which continues to be updated from the first step. Within each device ($i$), there are $\mathcal {O}(\log (n))$ cached models. Each model corresponds to the local model of a device ($j$) such that Device $i$ is a neighbor of Device $j$, i.e., $\omega_{i,j} > 0$. When Device $j$ sends its model ($m_j^{t_j}$) to Device $i$, the corresponding cache $j$ on Device $i$ is updated. This process is atomic to ensure that the whole model is available for model aggregation. As a model may take much memory space, e.g., LLaMA \cite{touvron2023llama}, the cached model can be either placed on the memory of the GPU (GPU RAM) or the memory of the device (CPU RAM). When the model is placed on the memory of the device, 
it may take extra time to move the model to the GPU within the model aggregation process. Thus, we first choose the GPU memory by default. When the GPU memory is not enough or there are numerous devices in the system, we take extra memory space of the device for the cached models. The consensus distance can be calculated based on the cached models and the local model. 
\vspace{-4mm}
\section{Dynamic Model Aggregation}
\label{sec:modelAggre}

In this section, we propose our dynamic model aggregation method. First, we propose a novel reinforcement learning-based model selection method. Then, we present a dynamic weight adjustment strategy. Last but not least, we present a model aggregation method to merge diverse models. 


In order to achieve the objective defined in Formula \ref{eq:problem}, we dynamically aggregate the local model and the cached neighbor models. We formulate the problem of the local update process as a bi-level optimization problem \cite{Bard1998PracticalBO} as defined in Formula \ref{eq:localProblem}: 
\vspace{-1mm}
\begin{align}
\min_{m_i, \omega_i} &\left[F_i(m_i, \omega_i)\triangleq\frac{1}{s_i}\sum_{(x_i, y_i) \in \mathcal{D}_i} f(agg(m_i, \omega_i), x_i, y_i)\right], \nonumber \\
s.t.~&\sum \omega_i = 1,
\label{eq:localProblem}
\end{align}
where $agg(m_i, \omega_i)$ represents the model aggregation process, $\omega_i = \{\omega_{i,i}, \omega_{i,i + 1}, \omega_{i,i + 2}, ..., \omega_{i,i + 2^m}\}$ represents a set of weights for the local model and neighbor models within the model aggregation process and the sum equals to 1. 



\vspace{-4mm}
\subsection{Model Selection}

In this section, we propose a reinforcement learning-based method to select proper neighbor models for model aggregation. While the training process is asynchronous, the cached models within each device are of diverse versions. For instance, Device $i$ is among the neighbor devices of Devices $j$ and $j'$. 
Then, the cached models for Devices $j$ and $j'$ are $m_j^{t_j}$ and $m_{j'}^{t_{j'}}$ with $t_j$ and $t_{j'}$ representing the number local updates executed on Device $i$ when receiving $m_j^{t_j}$ and $m_{j'}^{t_{j'}}$, respectively. We denote the number of local updates on Device $i$ by $t_i$. Then, when $t_j << t_i$, the cached model $m_j^{t_j}$ may not be beneficial within the model aggregation because of stale knowledge. When $t_{j'} \geq t_i$, it is of much possibility to improve the local model of Device $i$ with 
$m_{j'}^{t_{j'}}$. 
In addition, some models may be already aggregated in a previous local update, which can be ignored in the following model aggregation. Thus, we construct a reinforcement learning-based model to intelligently select proper neighbor models for the model aggregation process.

The reinforcement learning-based model consists of two modules. The first module is a priority neural network composed of a Long Short-Term Memory network and two fully connected layers. The output of the priority neural network is the priority possibility to choose each neighbor model. The second module is a priority converter, which selects the neighbor models based on the priority possibility. 

Within the training process on a device, we define the reward as the loss value calculated based on the local loss function, i.e., $\mathcal{F}_{i}(\boldsymbol{w}_{i}^{t_i}, \zeta_{i}^{t_i})$ with $\zeta_{i}^{t_i}$ representing the sampled data in SGD. Inspired by \cite{williams1992simple,Zoph2017Neural}, we update the priority neural network based on Formula \ref{eq:rlupdate} as follows:
\vspace{-1mm}
\begin{equation}\label{eq:rlupdate}
    \theta_{t_i+1} = \theta_{t_i} - \eta' \sum_{m = 0}^{ \log (n)} \nabla_{\theta_{t_i}} \log P(c_m|c_{(m-1):1}; \theta_{t})(\mathcal{R}_{t} - l_i),
    \vspace{-2mm}
\end{equation}
where $\theta_{t_i}$ represents the parameters in the priority neural network on Device $i$ at $t$-th local training, $\eta'$ refers to the learning rate, $\log (n)$ is the number of cashed neighbor models, $c_m$ corresponds to whether Model $m$ is selected, current local loss ($\mathcal{R}_{t}$) on device $i$ is the reward, i.e., $\mathcal{R}_{t}=\mathcal{F}_{i}(\boldsymbol{w}_{i}^{t_i}, \zeta_{i}^{t_i})$, and $l_i$ is a constant value for the bias, i.e., the average loss of the last certain times of local training on Device $i$. The input of the model includes whether the model is aggregated, the staleness of the model, and the loss value of the model. The model can be pre-trained with some heuristics, e.g., the model that is already aggregated should not be selected for the following model aggregation (which is synthetic model selection data), which correspond to the profiling results from real training process. Then, the model can be updated during the training process of \TheName{}. 
To the best of our knowledge, we are among the first to propose a reinforcement learning-based approach to select the model of neighbors for model aggregation so as to improve accuracy.

\vspace{-4mm}
\subsection{Dynamic Weight Update}
\label{subsec:dynamicWeight}

In this section, we propose our dynamic weight update method. In order to address the bi-level optimization problem defined in Formula \ref{eq:localProblem}, we propose a dynamic weight update method for the model aggregation process. We exploit the SGD method to update the local model $m_i$ as defined in Formula \ref{eq:localTraining} for the model aggregation on Device $i$. 
\vspace{-2mm}
\begin{equation}
\label{eq:localTraining}
    m^{t_i+\frac{1}{2}}_i \gets m^{t_i}_i - \eta^{t_i} \nabla_{m^{t_i}_i} F_i(m^{t_i}_i,\omega^{t_i}_i),
\end{equation}
where $\nabla_{m^{t_i}_i} F_i(m^{t_i}_i,\omega^{t_i}_i)$ represents the gradients in SGD, and $\eta^{t_i}$ refers to the learning rate. Then, we dynamically update the weights of neighbor models $\omega_i$. We use Formula \ref{eq:importanceDef} to calculate the importance of neighbor model $j$:
\vspace{-2mm}
\begin{equation}
\label{eq:importanceDef}
    \omega^{'t_i}_{i,j} = \frac{s_j * \lambda^{t_i}_{i,j}}{\sqrt{\bigtriangleup t^{t_i}_{i,j}} * loss^{t_i}_j},
\vspace{-2mm}
\end{equation}

where $\omega^{'t_i}_{i,j}$ is the importance of the cached model from Device $j$ on Device $i$, $\lambda^{t_i}_{i,j}$ is a control parameter to be dynamically updated, $\bigtriangleup t^{t_i}_{i,j}$ represents the difference between the current number of local update $t_i$ and the number of local updates when the model is updated from Device $j$, $loss^{t_i}_j$ refers to the loss of the model on Device $j$. $\bigtriangleup t^{t_i}_{i,j}$ can well represent the staleness of the neighbor model of Device $j$ on Device $i$. While a doubly-stochastic weight matrix can help obtain a consensual solution \cite{yuan2016convergence,shi2015extra}
we calculate the weight of neighbor model $j$ with Formula \ref{eq:weightDef}.
\vspace{-3mm}
\begin{equation}
\label{eq:weightDef}
    \omega^{t_i}_{i,j} = \frac{\omega^{'t_i}_{i,j}}{\sum_{j = i\text{ or }j \in \mathcal{M}} \omega^{'t_i}_{i,j}},
    \vspace{-3mm}
\end{equation}
where $\omega^{t_i}_{i,j}$ is the weight of the cached model from Device $j$ on Device $i$ in $t_i$-th local update, and $\mathcal{M}$ refers to the set of neighbors of Device $i$. In order to minimize the loss value, we update the control parameter $\lambda^{t_i}_{i,j}$ exploiting formula \ref{eq:control}.
\vspace{-5mm}
\begin{align}
\label{eq:control}
\lambda^{t_i}_{i,j} &= \lambda^{t_i-1}_{i,j} - \eta_{\lambda}\nabla_{\lambda^{t_i-1}_{i,j}} F_i(m^{t_i}_i), 
\end{align}
where $\eta_{\lambda}$ is the learning rate, and $m^{t_i}_i$ is calculated based on Formula \ref{eq:aggregate}:
\vspace{-4mm}
\begin{equation}
\label{eq:aggregate}
    m^{t_i}_i = \sum_{j = i\text{ or }j \in \mathcal{M}} \omega^{t_i - 1}_{i,j} m^{t_i - 1}_j.
\end{equation}
The partial derivatives of the loss function on the control parameter are calculated based on Formula \ref{eq:derivative}:
\vspace{-3mm}
\begin{align}
\label{eq:derivative}
    &\nabla_{\lambda^{t_i-1}_{i,j}} F_i(m^{t_i}_i) \\
    = &~\frac{\sum_{k = i\text{ or }(k \in \mathcal{M}\text{ and }k \neq j)} \omega^{'t_i-1}_{i,k}}{(\sum_{k = i\text{ or }k \in \mathcal{M}} \omega^{'t_i-1}_{i,k})^2} \frac{s_j g_i^\mathrm{T} m^{t_i - 1}_j}{\sqrt{\bigtriangleup t^{t_i-1}_{i,j}} * loss^{t_i-1}_j},\nonumber
\end{align}
where $g_i$ represents the gradients of $m^{t_i}_i$ on Device $i$.

\subsection{Heterogeneous Model Aggregation}
\label{subsec:heteModelAggre}

In this section, we present our heterogeneous model aggregation method with the updated weights. Because of the pruning process (see details in Section \ref{sec:sparseTraining}), the cached neighbor models may be heterogeneous in terms of structure. We exploit a full model structure as shown in Figure \ref{fig:localUpdate} and masks for each neighbor model and local model for the aggregation. A full model structure is an original model without pruning. The mask is utilized to identify which 
\begin{figure}[t]
\vspace{-4mm}
\begin{algorithm}[H]
\caption{Dynamic Model Aggregation}
\label{alg:dmaAlgo}
\begin{algorithmic}[1]
\REQUIRE  \quad \newline
$i$: The index of the device \newline
$m^0_i$: The initial model on Device $i$ \newline
$T$: The maximum number of local updates \newline
$M$: The set of neighbor models 
\ENSURE \quad \newline
$m^{t_i}_i$: The global model at Round $t_i$ 
\FOR{$t_i$ in $\{1, 2, ..., T\}$}
\STATE $m^{t_i-\frac{1}{2}}_i \gets $ update $m^{t_i-1}_i$ based on Formula \ref{eq:localTraining} \label{line:SGD}
\FOR{$j \in \mathcal{M}$}
\STATE Update $\lambda^{t_i}_{i,j}$ according to Formula \ref{eq:control} \label{line:lambdaUpdate}
\STATE Update $\omega^{t_i}_{i,j}$ according to Formulas \ref{eq:importanceDef} and \ref{eq:weightDef} \label{line:weightUpdate}
\ENDFOR
\STATE $m^{t_i}_i \gets$ update $m^{t_i-\frac{1}{2}}_i$ with $M$ according to Formula \ref{eq:parameterAggregation} \label{line:modelAggregation}
\ENDFOR
\end{algorithmic}
\end{algorithm}
\vspace{-16mm}
\end{figure}
\vspace{-2mm}
neuron is kept and which neuron is removed after pruning. Then, for each parameter ($\mu^p$) in the full model, we can calculate its aggregated value based on Formula \ref{eq:parameterAggregation}.

\begin{equation}
\label{eq:parameterAggregation}
    \mu^p_i = \frac{1}{\sum_{j = i\text{ or }j \in \mathcal{M}} \omega^{t_i - 1}_{i,j} o^p_j}\sum_{j = i\text{ or }j \in \mathcal{M}} \omega^{t_i - 1}_{i,j} \mu^p_{i,j} o^p_j,
    \vspace{-6mm}
\end{equation}
where $o^p_j$ is the mask of the model from neighbor Device $j$ with 1 representing the corresponding parameter $\mu^p$ is not pruned and 0 representing that $\mu^p$ is pruned. After calculating the parameter values in the full model, the local model can be generated by exploiting the local mask. 

As shown in Algorithm \ref{alg:dmaAlgo}, the dynamic model aggregation algorithm consists of multiple local updates. Within each local update, the local model is updated with the SGD method (Line 2). Then, for each model in $\mathcal{M}$, the control parameter is updated (Line 4) and the corresponding weight value is updated (Line 5). Afterward, the local model and the neighbor models are aggregated based on Formula \ref{eq:parameterAggregation} (Line 7). The convergence analysis can be achieved based on the existing theoretical works \cite{yuan2016convergence,
lian2018asynchronous}, which is out of the scope of this paper.


\vspace{-4mm}
\section{Sparse Training}
\label{sec:sparseTraining}

We propose a novel adaptive sparse training method to reduce computation and communication costs while minimizing the impact on the loss function and considering the exploration of the sensitivity of neurons, i.e., the impact of the pruned parameters in the next rounds of updates. We assume that at specific rounds of local training, e.g., $t_i$, we carry out the pruning operation on the whole model, i.e., the model with all the parameters. Within the pruning operation, we denote the modification of the local model $m_i$ on Device $i$ by $\Delta m_i$. Then, we denote the impact on the current local loss function by $\Delta F_i^C$, which can be calculated via Taylor expansion as defined in Formula \ref{eq:deltaLoss}.
\begin{align}
\label{eq:deltaLoss}
    \Delta F_i^C = &~F_i(m_i + \Delta m_i) - F_i(m_i) \nonumber\\
               = &~g^T_i \Delta m_i + \frac{1}{2} \Delta m^{T}_i \mathcal{H} \Delta m_i + \mathcal{O}(||\Delta m_i||^3),
\end{align}
where $g_i$ is the gradient, $\mathcal{H}$ represents the Hessian matrix of the model, and $\mathcal{O}(||\Delta m_i||^3)$ corresponds to higher-order items, which can be ignored \cite{hassibi1993optimal,yu2022hessian}. We retrieve the Hessian matrix efficiently utilizing the PyHessian library \cite{yao2020pyhessian} with small computation and storage costs. We associate the pruned parameters (channels) to $p$ and the remained parameters to $r$. Then, we have:
\vspace{-2mm}
\begin{align}
\label{eq:LossExp}
    \Delta F_i^C = &~g^T_i \Delta m_i + \frac{1}{2} \Delta m^{T}_i \mathcal{H} \Delta m_i \nonumber\\
    = &~\begin{pmatrix} g_i^p \\ g_i^r \end{pmatrix} ^T \begin{pmatrix} \Delta m_i^p \\ \Delta m_i^r \end{pmatrix} \\
    &~ + \frac{1}{2} \begin{pmatrix} \Delta m_i^p \\ \Delta m_i^r \end{pmatrix}^T \begin{pmatrix} \mathcal{H}^{p,p},\mathcal{H}^{p,r} \\  \mathcal{H}^{r,p},\mathcal{H}^{r,r} \end{pmatrix} \begin{pmatrix} \Delta m_i^p \\ \Delta m_i^r \end{pmatrix}. \nonumber
\end{align}



Please note that the pruning operation is carried out during the training process. Thus, we cannot ignore $g_i$, i.e., $g^T_i \neq 0$. 
By combining Formula \ref{eq:mi} into Formula \ref{eq:LossExp}, we can have Formula \ref{eq:LossLagrange} minimizing the current impact $\Delta F_i^C$.

\vspace{-5mm}
\begin{align}
\label{eq:LossLagrange}
    \Delta F_i^C =&~\frac{1}{2} (m_i^p)^T \mathcal{H}^{p,p} m_i^p - (g_i^p)^T m_i^p\nonumber\\
    &~- \frac{1}{2}(m_i^p)^T \mathcal{H}^{p,r} (\mathcal{H}^{r,r})^{-1} \mathcal{H}^{r,p} m_i^p \nonumber\\
    &~- \frac{1}{2}(g_i^r)^T (\mathcal{H}^{r,r})^{-1} g_i^r \nonumber\\
    &~+ (g_i^r)^T (\mathcal{H}^{r,r})^{-1} \mathcal{H}^{r,p} m_i^p. 
\end{align}
However, Formula \ref{eq:deltaLoss} only considers the impact of the current loss value. We further consider the exploration of the pruned parameters by inserting the magnitude of the gradients as Defined in Formula \ref{eq:deltaLossTotal}.
\vspace{-2mm}
\begin{align}
\label{eq:deltaLossTotal}
    \Delta F_i = &~(1-\lambda_g)\frac{|\Delta F_i^C|}{|F_i|} + \lambda_g \frac{||\Delta g_i||_2}{||g_i||_2},
\end{align}
where $|\cdot|$ represents the absolute value and $0 \leq \lambda_g \leq 1$ is a hyper-parameter. The added term $\lambda_g \frac{||\Delta g_i||_2}{||g_i||_2}$ represents the exploration of the training process, which should be significant at the beginning of the training process and small at the end. When the magnitude of $g_i$ is significant, the model is not well trained and we should pay attention to the exploration. We empirically set 
$\lambda_g = \frac{||g_i||_2}{\mathcal{C}||g_i||_2^{max}}$, with $\mathcal{C}$ being a hyper-parameter, $g_i$ representing the current gradients and $||g_i||_2^{max}$ is the maximum $L^2$ norm of the gradients in the previous training process. 
The setting of $\lambda_g$ can help reduce the impact of pruned parameters on gradients at the beginning and that on loss at the end so as to improve the accuracy of the model, as the gradients are important to improve the model at the beginning. In practice, we fine-tune the value of $\mathcal{C}$ in the experimentation. 
As presented in Section \ref{subsec:heteModelAggre}, for a device, although the pruned model is transferred and updated in the local update, the pruned parameters are still updated when they remain in neighbor models. Then, during each pruning process, we consider the whole model as the original model to preserve the parameters that potentially become important.

Furthermore, as pruning rates are critical to the pruning process \cite{zhang2021unified}, we automatically generate a proper pruning rate for the pruning operation. With the recent success of lottery ticket for model pruning \cite{babakniya2022federated}, we exploit a lossless method to calculate the pruning rate. On Device $i$, we denote the initial model by $m^o_i$ and the difference between the current model and the initial model by $\Delta m_i = m_i^o - m_i$. We sort the eigenvalues of the Hessian matrix $\mathcal{H}(m_i)$ in ascending order, i.e., $\{h^p_i|p \in (1, d_i)\}$ with $d_i$ referring to the rank of the Hessian matrix and $p$ representing the index. Then, we calculate the Lipschitz constant, denoted as $\mathscr{L}_i$, of a benchmark function $F'_i(\Delta m_i) = \mathcal{H}(m_i) - \triangledown F_i(\Delta m_i + m_i)$. We take the first eigenvalue $h^p_i$ that satisfies $h^{p+1}_i - h^p_i > 4\mathscr{L}_i$ to calculate a proper pruning rate by $p^*_i = \frac{p}{d_{k}}$ to achieve lossless pruning \cite{zhang2021validating}. Afterward, we calculate the pruning rate with Formula \ref{eq:pruningRateAggregation} with the consideration of neighbors.
\vspace{-2mm}
\begin{equation}
\label{eq:pruningRateAggregation}
    p_i = \sum_{j \in \mathcal{M}} \omega^{t_i}_{i,j} p_j + \omega^{t_i}_{i,i}p^*_i,
\vspace{-2mm}
\end{equation}
where $\omega^{t_i}_{i,j}$ refers to the weights calculated in Section \ref{subsec:dynamicWeight}. Please see the detailed algorithm for our adaptive pruning method in Section \ref{sec:adaptive_pruning} in the Appendix.


\vspace{-4mm}
\section{Experiments}
\label{sec:exp}

In this section, we demonstrate evaluation results for \TheName{}. First, we present the experimental setup. Then, we show the comparison of \TheName{} with 14 baseline approaches. Afterward, we explain the advantages of dynamic model aggregation and the sparse training with the ablation study.

\subsection{Setup}

\vspace{-4mm}
We consider a decentralized FL system with 100 devices with an exponential graph topology. We exploit 3 datasets, i.e., Emnist-letters (Emnist) \cite{cohen2017emnist}, CIFAR-10 (CIFAR) \cite{krizhevsky2009learning}, Tiny-ImageNet \cite{le2015tiny}, and 3 models, i.e., LeNet-5 (LeNet) \cite{lecun1998gradient}, Resnet-8 (ResNet) \cite{He2016}, VGG-9 (VGG) \cite{simonyan2015very}, for image classification tasks. We further conduct experiments on a Natural Language Processing (NLP) task, i.e., sentiment analysis with the IMDb dataset on TextCNN \cite{zhou2021distilled}, to demonstrate the generality of \TheName{}. We utilize the Dirichlet distribution \cite{li2021federated} to partition the data. Please see details in the Appendix. 

\begin{table*}
  \scriptsize
  \caption{The accuracy, training time, and computation costs with \TheName{} and diverse baseline approaches. ``Acc'' represents the convergence accuracy. ``Time'' refers to the training time (s) to achieve a target accuracy, i.e., 0.9 for LeNet with Emnist, 0.6  for ResNet with CIFAR, and 0.17 for VGG with Tiny-ImageNet, 0.17 for ResNet with Tiny-ImageNet, and 0.65 for IMDb with TextCNN. 
  ``MFP'' represents the computational costs (MFLOPs). ``/'' represents that training does not achieve the target accuracy.
  }
  \label{tab:cmp_ASDFL}
  \centering
  \begin{tabular}{l|lll|lll|lll|lll|lll}
    \toprule
    \multirow{2}{*}{Method} & \multicolumn{3}{c|}{Emnist \& LeNet}  & \multicolumn{3}{c|}{CIFAR \& ResNet} & \multicolumn{3}{c|}{TinyImageNet \& VGG} &  \multicolumn{3}{c|}{TinyImageNet \& ResNet} &  \multicolumn{3}{c}{IMDb \& TextCNN}\\
    \cmidrule(r){2-16}  & Acc       & Time   & MFP    & Acc       & Time   & MFP & Acc       & Time   & MFP    & Acc     & Time      & MFP  & Acc     & Time      & MFP \\
    \midrule
    \TheName{}          &  \textbf{0.9326}    & \textbf{384}   &  \textbf{0.172}      & \textbf{0.7453} &  \textbf{314}    & \textbf{1.29}   &  \textbf{0.2025}   & \textbf{884}  & \textbf{61.8}   &  \textbf{0.2696}  & \textbf{1085}  & \textbf{368}    & \textbf{0.7976} & \textbf{517}   & \textbf{0.389}\\
    FedAvg              &  0.9159    & 817   &  0.283       & 0.6600    &  3133     & 1.74     &  0.1713    & 5360   &  73.6       &  0.2254    & 3776    &  452   & 0.7763  & 1329  & 0.652   \\
    FedProx             &  0.9174    & 696   &  0.283       & 0.6136    &  3935     & 1.74     &  0.1726    & 4357   &  73.6       &  0.1645    & /       &  452   & 0.7625  & 1289  & 0.652   \\
    FedNova             &  0.9160    & 998   &  0.283       & 0.6493    &  3796     & 1.74     &  0.1700    & 4564   &  73.6       &  0.2312    & 3567    &  452   & 0.7737  & 1269  & 0.652   \\
    SAFA                &  0.9173    & 800   &  0.283       & 0.6751    &  2050     & 1.74     &  0.1701    & 1742   &  73.6       &  0.2305    & 10101   &  452   & 0.7488  & 1243  & 0.652\\
    Sageflow            &  0.9168    & 767   &  0.283       & 0.6755    &  1614     & 1.74     &  0.1740    & 1742   &  73.6       &  0.2323    & 2863    &  452   & 0.7529  & 1160  & 0.652\\
    FedSA               &  0.9158    & 397   &  0.283       & 0.6613    &  3090     & 1.74     &  0.1539    & 1658   &  73.6       &  0.2123    & 1931    &  452   & 0.7746  & 1105  & 0.652\\
    ASO-Fed             &  0.9136    & 600   &  0.283       & 0.6652    &  1030     & 1.74     &  0.1529    & 1419   &  73.6       &  0.2229    & 1707    &  452   & 0.7814  & 1092  & 0.652\\
    Port                &  0.9165    & 419   &  0.283       & 0.6596    &  1243     & 1.74     &  0.1477    & 1632   &  73.6       &  0.2155    & 2149    &  452   & 0.7620  & 1188  & 0.652\\
    FedBuff             &  0.9177    & 497   &  0.283       & 0.6647    &  3379     & 1.74     &  0.1487    & 1465   &  73.6       &  0.2118    & 1879    &  452   & 0.7890  & 977  & 0.652\\
    AD-PSGD             &  0.9129    & 940   &  0.283       & 0.6384    &  4346     & 1.74     &  0.1165    & /      &  73.6       &  0.1938    & 1438    &  452   & 0.7724  & 552  & 0.652\\
    DisPFL              &  0.9175    & 402   &  0.175       & 0.6473    &  767      & 1.48     &  0.1530    & 1067   &  62.5       &  0.1852    & 5350    &  384   & 0.7448  & 764  & 0.587\\
    Hrank               &  0.9270    & 394   &  0.277       & 0.7121    &  545      & 1.46     &  0.1849    & 975    &  62.5       &  0.2391    & 2636    &  384   & 0.7675  & 683  & 0.587\\
    FedAP               &  0.9280    & 417   &  0.278       & 0.7189    &  647      & 1.39     &  0.1806    & 936    &  63.5       &  0.2381    & 2991    &  383   & 0.7504  & 1454  & 0.463\\
    HAP                 &  0.8752    & 402   &  0.173       & 0.6319    &  572      & 1.49     &  0.1632    & 1010   &  62.6       &  0.1794    & 3947    &  384   & 0.7668  & 928  & 0.587\\
    \bottomrule
  \end{tabular}
  \vspace{-5mm}
\end{table*}


We take 14 existing approaches as baselines, i.e., FedAvg \cite{mcmahan2017communication}, FedProx \cite{Li2020FedProx}, FedNova \cite{Wang2020Tackling}, SAFA \cite{wu2020safa}, Sageflow \cite{park2021sageflow}, AD-PSGD \cite{lian2018asynchronous}, FedSA \cite{chen2021fedsa}, ASO-Fed \cite{Chen2020Asynchronous}, FedBuff \cite{nguyen2022federated}, Port \cite{Su2022How}, Hrank \cite{lin2020hrank}, FedAP \cite{Zhang2022FedDUAP}, HAP \cite{yu2022hessian}, DisPFL \cite{Dai2022DisPFLTC}. In addition, we denote the version with Reinforcement Learning-based model selection by \TheName{}-RL, the version with Dynamic Weight Update by \TheName{}-DWU, and the version with sparse training by \TheName{}-P, the version with dynamic model aggregation by \TheName{}-RL-DWU, the version with all three modules by \TheName{} and the version without any module by \TheName{}-0.

\vspace{-4mm}
\subsection{Evaluation of \TheName{}}
\label{subsec:evaluation}

As shown in Table \ref{tab:cmp_ASDFL}, \TheName{} corresponds to the highest accuracy and the fastest training speed among all the approaches. 
\TheName{} significantly outperforms other baseline approaches in the whole training process with a small dataset (Emnist) and model (LeNet). The advantage of \TheName{} can be up to 5.8\% in terms of convergence accuracy, 61.5\% in terms of the training time, and 39.1\% in terms of computation costs. 

When the dataset and the model become complicated, e.g., CIFAR with ResNet or VGG and Tiny-ImageNet with ResNet or VGG, \TheName{} significantly outperforms other baseline approaches in terms of the convergence accuracy. 
At the beginning of the training, \TheName{} performs the dynamic adjustment of the weights for neighbor models, which can bring high accuracy in the middle or at the end of the training process. As shown in Table \ref{tab:cmp_ASDFL}, the advantages of \TheName{} can be up to 8.5\% for FedAvg, 13.2\% for FedProx, 9.6\% for FedNova,  10\% for SAFA, 7\% for Sageflow, 8.4\% for FedSA, 8\% for ASO-Fed, 8.6\% for Port, 8.9\% for FedBuff, 15.4\% for AD-PSGD, 9.8\% for DisPFL, 7.7\% for Hrank 7.3\% for FedAP, and 11.3\% for HAP in terms of accuracy. Furthermore, \TheName{} corresponds to a short training time. The training time of \TheName{} to achieve a target accuracy can be up to 90\% (compared with FedAvg), 92\% (compared with FedProx), 91.7\% (compared with FedNova), 84.7\% (compared with SAFA), 82.2\% (compared with Sageflow), 89.8\% (compared with FedSA), 69.5\% (compared with ASO-Fed), 84.9\% (compared with Port), 90.7\% (compared with FedBuff), 92.8\% (compared with AD-PSGD), 71\% (compared with DisPFL), 52.2\% (compared with Hrank), 56.3\% (compared with FedAP), and 62.8\% (compared with HAP) shorter. 
Finally, as the sparse training can well reduce the model size, \TheName{} corresponds to much smaller computation costs (up to 42.3\% compared with FedAvg, FedProx, FedNova, SAFA, Sageflow, FedSA, ASO-Fed, Port, FedBuff, and AD-PSGD, 12.5\% compared with DisPFL, 11.5\% compared with Hrank, 6.9\% compared with FedAP, and 13.6\% compared with HAP), while the corresponding accuracy remains the highest with the shortest time to achieve target accuracy. 
Meanwhile, the minimal advantages of our approach are significant as well with the most complicated setting (Tiny-ImageNet with ResNet), i.e., 3.1\% higher accuracy, 24.5\% faster to achieve target accuracy, and 3.9\% smaller computational cost.



We further conduct sentiment analysis experiments on the IMDb dataset with TextCNN to demonstrate the generality of \TheName{}. As shown in Table \ref{tab:cmp_ASDFL}, the advantages of \TheName{} can be up to 5.3\% in accuracy, 64.4\% in training speed and 40.3\% in computation costs.
The results reveal that \TheName{} can be easily adopted across various tasks.

Furthermore, we carry out the experimentation with different network bandwidth, multiple values of $\mathcal{C}$ (please refer to Section \ref{sec:sparseTraining}), and diverse device heterogeneity in the Appendix.

\vspace{-4mm}



\section{Conclusion}
\label{sec:con}

In this paper, we propose a novel \underline{A}synchronous \underline{E}fficient \underline{D}ecentralized \underline{F}ederated \underline{L}earning framework (\TheName{}), with three original contributions, i.e., an asynchronous decentralized FL system model, a dynamic model aggregation method consisting of a reinforcement learning-based model selection method and a dynamic staleness-based weight update strategy, and an adaptive pruning method for sparse training. We carry out extensive experiments with four models and four public datasets to demonstrate the significant advantages of \TheName{} in terms of accuracy (up to 16.3\% higher), efficiency (up to 92.9\% faster), and computation costs (up to 42.3\% smaller).

\vspace{1mm}
{\small
\bibliographystyle{siam}
\bibliography{ref}

\begin{thebibliography}{10}

\bibitem{adlam2020understanding}
{\sc B.~Adlam and J.~Pennington}, {\em Understanding double descent requires a
  fine-grained bias-variance decomposition}, NeurIPS, 33 (2020),
  pp.~11022--11032.

\bibitem{assran2019stochastic}
{\sc M.~Assran, N.~Loizou, N.~Ballas, and M.~Rabbat}, {\em Stochastic gradient
  push for distributed deep learning}, in ICML, 2019, pp.~344--353.

\bibitem{babakniya2022federated}
{\sc S.~Babakniya, S.~Kundu, S.~Prakash, Y.~Niu, and S.~Avestimehr}, {\em
  Federated sparse training: Lottery aware model compression for resource
  constrained edge}, in FL-NeurIPS, 2022, pp.~1--11.

\bibitem{Bard1998PracticalBO}
{\sc J.~F. Bard}, {\em Practical Bilevel Optimization: Algorithms and
  Applications}, Springer, New York, USA, 1998.

\bibitem{bibikar2022federated}
{\sc S.~Bibikar, H.~Vikalo, Z.~Wang, and X.~Chen}, {\em Federated dynamic
  sparse training: Computing less, communicating less, yet learning better}, in
  AAAI, 2022, pp.~6080--6088.

\bibitem{bonawitz2019towards}
{\sc K.~Bonawitz, H.~Eichner, W.~Grieskamp, D.~Huba, A.~Ingerman, V.~Ivanov,
  C.~Kiddon, et~al.}, {\em Towards federated learning at scale: System design},
  MLSys,  (2019), pp.~374--388.

\bibitem{che2023federated}
{\sc T.~Che, J.~Liu, Y.~Zhou, J.~Ren, J.~Zhou, V.~S. Sheng, H.~Dai, and
  D.~Dou}, {\em Federated learning of large language models with
  parameter-efficient prompt tuning and adaptive optimization}, in {EMNLP},
  2024, pp.~1--18.

\bibitem{che2022federated}
{\sc T.~Che, Z.~Zhang, Y.~Zhou, X.~Zhao, J.~Liu, Z.~Jiang, D.~Yan, R.~Jin, and
  D.~Dou}, {\em Federated fingerprint learning with heterogeneous
  architectures}, in ICDM, IEEE, 2022, pp.~31--40.

\bibitem{che2023fast}
{\sc T.~Che, Y.~Zhou, Z.~Zhang, L.~Lyu, J.~Liu, D.~Yan, D.~Dou, and J.~Huan},
  {\em Fast federated machine unlearning with nonlinear functional theory}, in
  ICML, PMLR, 2023, pp.~4241--4268.

\bibitem{chen2021fedsa}
{\sc M.~Chen, B.~Mao, and T.~Ma}, {\em Fedsa: A staleness-aware asynchronous
  federated learning algorithm with non-iid data}, FGCS, 120 (2021), pp.~1--12.

\bibitem{Chen2020Asynchronous}
{\sc Y.~Chen, Y.~Ning, M.~Slawski, and H.~Rangwala}, {\em Asynchronous online
  federated learning for edge devices with non-iid data}, in Big Data, {IEEE},
  2020, pp.~15--24.

\bibitem{cohen2017emnist}
{\sc G.~Cohen, S.~Afshar, J.~Tapson, and A.~Van~Schaik}, {\em Emnist: Extending
  mnist to handwritten letters}, in IJCNN, {IEEE}, 2017, pp.~2921--2926.

\bibitem{Dai2022DisPFLTC}
{\sc R.~Dai, L.~Shen, F.~He, X.~Tian, and D.~Tao}, {\em Dispfl: Towards
  communication-efficient personalized federated learning via decentralized
  sparse training}, in ICML, 2022, pp.~1--18.

\bibitem{d2020double}
{\sc S.~d’Ascoli, M.~Refinetti, G.~Biroli, and F.~Krzakala}, {\em Double
  trouble in double descent: Bias and variance (s) in the lazy regime}, in
  ICML, 2020, pp.~2280--2290.

\bibitem{hassibi1993optimal}
{\sc B.~Hassibi, D.~G. Stork, and G.~J. Wolff}, {\em Optimal brain surgeon and
  general network pruning}, in ICNN, 1993, pp.~293--299.

\bibitem{He2016}
{\sc K.~He, X.~Zhang, S.~Ren, and J.~Sun}, {\em Deep residual learning for
  image recognition}, in CVPR, 2016, pp.~770--778.

\bibitem{hegedHus2019decentralized}
{\sc I.~Heged{\H{u}}s, G.~Danner, and M.~Jelasity}, {\em Decentralized
  recommendation based on matrix factorization: a comparison of gossip and
  federated learning}, in ECML PKDD, W\"urzburg, Germany, 2019, Springer,
  pp.~317--332.

\bibitem{hsu2020federated}
{\sc T.-M.~H. Hsu, H.~Qi, and M.~Brown}, {\em Federated visual classification
  with real-world data distribution}, in ECCV, Glasgow, UK, 2020, Springer,
  pp.~76--92.

\bibitem{huang2022achieving}
{\sc T.~Huang, S.~Liu, L.~Shen, F.~He, W.~Lin, and D.~Tao}, {\em Achieving
  personalized federated learning with sparse local models}, arXiv:2201.11380,
  (2022), pp.~1--19.

\bibitem{huba2022papaya}
{\sc D.~Huba, J.~Nguyen, K.~Malik, R.~Zhu, M.~Rabbat, A.~Yousefpour, C.-J. Wu,
  H.~Zhan, P.~Ustinov, H.~Srinivas, et~al.}, {\em Papaya: Practical, private,
  and scalable federated learning}, MLSys,  (2022), pp.~814--832.

\bibitem{jiang2019networked}
{\sc J.~Jiang, Y.~Zhou, G.~Ananthanarayanan, Y.~Shu, and A.~A. Chien}, {\em
  Networked cameras are the new big data clusters}, in HotEdgeVideo, 2019,
  pp.~1--7.

\bibitem{JinAccelerated2022}
{\sc J.~Jin, J.~Ren, Y.~Zhou, L.~Lv, J.~Liu, and D.~Dou}, {\em Accelerated
  federated learning with decoupled adaptive optimization}, in ICML, vol.~162,
  2022, pp.~10298--10322.

\bibitem{mcmahan2021advances}
{\sc P.~Kairouz, H.~B. McMahan, B.~Avent, A.~Bellet, and M.~B. et~al.}, {\em
  Advances and open problems in federated learning}, Foundations and
  Trends{\textregistered} in Machine Learning,  (2021), pp.~1--210.

\bibitem{Karimireddy2020SCAFFOLD}
{\sc S.~P. Karimireddy, S.~Kale, M.~Mohri, S.~Reddi, S.~Stich, and A.~T.
  Suresh}, {\em {SCAFFOLD}: Stochastic controlled averaging for federated
  learning}, in ICML, 2020, pp.~5132--5143.

\bibitem{krizhevsky2009learning}
{\sc A.~Krizhevsky, G.~Hinton, et~al.}, {\em Learning multiple layers of
  features from tiny images}, Technical report, University of Toronto, 1
  (2009), pp.~1--60.

\bibitem{lakshman2010cassandra}
{\sc A.~Lakshman and P.~Malik}, {\em Cassandra: a decentralized structured
  storage system}, ACM SIGOPS operating systems review, 44 (2010), pp.~35--40.

\bibitem{le2015tiny}
{\sc Y.~Le and X.~Yang}, {\em Tiny imagenet visual recognition challenge}, CS
  231N, 7 (2015), p.~3.

\bibitem{lecun1998gradient}
{\sc Y.~LeCun, L.~Bottou, Y.~Bengio, and P.~Haffner}, {\em Gradient-based
  learning applied to document recognition}, Proceedings of the IEEE, 86
  (1998), pp.~2278--2324.

\bibitem{Li2022FedHiSyn}
{\sc G.~Li, Y.~Hu, M.~Zhang, J.~Liu, Q.~Yin, Y.~Peng, and D.~Dou}, {\em
  Fedhisyn: A hierarchical synchronous federated learning framework for
  resource and data heterogeneity}, in ICPP, 2022, pp.~1--10.
\newblock To appear.

\bibitem{li2021blockchain}
{\sc J.~Li, Y.~Shao, K.~Wei, M.~Ding, C.~Ma, L.~Shi, Z.~Han, and H.~V. Poor},
  {\em Blockchain assisted decentralized federated learning (blade-fl):
  Performance analysis and resource allocation}, TPDS,  (2021), pp.~2401--2415.

\bibitem{li2021federated}
{\sc Q.~Li, Y.~Diao, Q.~Chen, and B.~He}, {\em Federated learning on non-iid
  data silos: An experimental study}, arXiv preprint arXiv:2102.02079,
  abs/2102.02079 (2021), pp.~1--20.

\bibitem{Li2020FedProx}
{\sc T.~Li, A.~K. Sahu, M.~Zaheer, M.~Sanjabi, A.~Talwalkar, and V.~Smith},
  {\em Federated optimization in heterogeneous networks}, in MLSys, 2020,
  pp.~429--450.

\bibitem{li2019privacy}
{\sc W.~Li, F.~Milletar{\`\i}, D.~Xu, N.~Rieke, J.~Hancox, W.~Zhu, M.~Baust,
  Y.~Cheng, S.~Ourselin, M.~J. Cardoso, et~al.}, {\em Privacy-preserving
  federated brain tumour segmentation}, in MLMI, 2019, pp.~133--141.

\bibitem{lian2017can}
{\sc X.~Lian, C.~Zhang, H.~Zhang, C.-J. Hsieh, W.~Zhang, and J.~Liu}, {\em Can
  decentralized algorithms outperform centralized algorithms? a case study for
  decentralized parallel stochastic gradient descent}, in NeurIPS, 2017,
  pp.~1--11.

\bibitem{lian2018asynchronous}
{\sc X.~Lian, W.~Zhang, C.~Zhang, and J.~Liu}, {\em Asynchronous decentralized
  parallel stochastic gradient descent}, in ICML, 2018, pp.~3043--3052.

\bibitem{lin2020hrank}
{\sc M.~Lin, R.~Ji, Y.~Wang, Y.~Zhang, B.~Zhang, Y.~Tian, and L.~Shao}, {\em
  Hrank: Filter pruning using high-rank feature map}, in CVPR, 2020,
  pp.~1529--1538.

\bibitem{liu2022distributed}
{\sc J.~Liu, J.~Huang, Y.~Zhou, X.~Li, S.~Ji, H.~Xiong, and D.~Dou}, {\em From
  distributed machine learning to federated learning: a survey}, KAIS, 64
  (2022), pp.~885--917.

\bibitem{Liu2024FedASMU}
{\sc J.~Liu, J.~Jia, T.~Che, C.~Huo, J.~Ren, Y.~Zhou, H.~Dai, and D.~Dou}, {\em
  Fedasmu: Efficient asynchronous federated learning with dynamic
  staleness-aware model update}, in AAAI, 2024, pp.~1--18.
\newblock To appear.

\bibitem{liu2022multi}
{\sc J.~Liu, J.~Jia, B.~Ma, C.~Zhou, J.~Zhou, Y.~Zhou, H.~Dai, and D.~Dou},
  {\em Multi-job intelligent scheduling with cross-device federated learning},
  TPDS, 34 (2022), pp.~535--551.

\bibitem{liu2023heterps}
{\sc J.~Liu, Z.~Wu, D.~Feng, M.~Zhang, X.~Wu, X.~Yao, D.~Yu, Y.~Ma, F.~Zhao,
  and D.~Dou}, {\em Heterps: Distributed deep learning with reinforcement
  learning based scheduling in heterogeneous environments}, FGCS,  (2023).

\bibitem{liu2023distributed}
{\sc J.~Liu, X.~Zhou, L.~Mo, S.~Ji, Y.~Liao, Z.~Li, Q.~Gu, and D.~Dou}, {\em
  Distributed and deep vertical federated learning with big data}, CCPE,
  (2023), p.~e7697.

\bibitem{liu2022asynchronous}
{\sc Q.~Liu, B.~Yang, Z.~Wang, D.~Zhu, X.~Wang, K.~Ma, and X.~Guan}, {\em
  Asynchronous decentralized federated learning for collaborative fault
  diagnosis of pv stations}, IEEE TNSE, 9 (2022), pp.~1680--1696.

\bibitem{liu2022decentralized}
{\sc W.~Liu, L.~Chen, and W.~Zhang}, {\em Decentralized federated learning:
  Balancing communication and computing costs}, IEEE TSIPN, 8 (2022),
  pp.~131--143.

\bibitem{mcmahan2017communication}
{\sc B.~McMahan, E.~Moore, D.~Ramage, S.~Hampson, and B.~A. y~Arcas}, {\em
  Communication-efficient learning of deep networks from decentralized data},
  in AISTATS, 2017, pp.~1273--1282.

\bibitem{nakkiran2021deep}
{\sc P.~Nakkiran, G.~Kaplun, Y.~Bansal, T.~Yang, B.~Barak, and I.~Sutskever},
  {\em Deep double descent: Where bigger models and more data hurt}, Journal of
  Statistical Mechanics: Theory and Experiment,  (2021), p.~124003.

\bibitem{nguyen2022federated}
{\sc D.~C. Nguyen, Q.-V. Pham, P.~N. Pathirana, M.~Ding, A.~Seneviratne,
  Z.~Lin, O.~Dobre, and W.-J. Hwang}, {\em Federated learning for smart
  healthcare: A survey}, CSUR,  (2022), pp.~1--37.

\bibitem{park2021sageflow}
{\sc J.~Park, D.-J. Han, M.~Choi, and J.~Moon}, {\em Sageflow: Robust federated
  learning against both stragglers and adversaries}, in NeurIPS, 2021,
  pp.~840--851.

\bibitem{qu2021decentralized}
{\sc Y.~Qu, H.~Dai, Y.~Zhuang, J.~Chen, C.~Dong, F.~Wu, and S.~Guo}, {\em
  Decentralized federated learning for uav networks: Architecture, challenges,
  and opportunities}, IEEE Network, 35 (2021), pp.~156--162.

\bibitem{shi2015extra}
{\sc W.~Shi, Q.~Ling, G.~Wu, and W.~Yin}, {\em Extra: An exact first-order
  algorithm for decentralized consensus optimization}, SIAM Journal on
  Optimization,  (2015), pp.~944--966.

\bibitem{simonyan2015very}
{\sc K.~Simonyan and A.~Zisserman}, {\em Very deep convolutional networks for
  large-scale image recognition}, in ICLR, San Diego, CA, USA, 2015, pp.~1--14.

\bibitem{Su2022How}
{\sc N.~Su and B.~Li}, {\em How asynchronous can federated learning be?}, in
  IWQoS, 2022, pp.~1--11.

\bibitem{touvron2023llama}
{\sc H.~Touvron, T.~Lavril, G.~Izacard, X.~Martinet, M.-A. Lachaux, T.~Lacroix,
  B.~Rozi{\`e}re, N.~Goyal, E.~Hambro, et~al.}, {\em Llama: Open and efficient
  foundation language models}, arXiv:2302.13971,  (2023), pp.~1--27.

\bibitem{vahidian2021personalized}
{\sc S.~Vahidian, M.~Morafah, and B.~Lin}, {\em Personalized federated learning
  by structured and unstructured pruning under data heterogeneity}, in ICDCSW,
  2021, pp.~27--34.

\bibitem{vanhaesebrouck2017decentralized}
{\sc P.~Vanhaesebrouck, A.~Bellet, and M.~Tommasi}, {\em Decentralized
  collaborative learning of personalized models over networks}, in AISTATS,
  2017, pp.~509--517.

\bibitem{Wang2020Tackling}
{\sc J.~Wang, Q.~Liu, H.~Liang, G.~Joshi, and H.~V. Poor}, {\em Tackling the
  objective inconsistency problem in heterogeneous federated optimization}, in
  NeurIPS, 2020, pp.~7611--7623.

\bibitem{williams1992simple}
{\sc R.~J. Williams}, {\em Simple statistical gradient-following algorithms for
  connectionist reinforcement learning}, Machine Learning, 8 (1992),
  pp.~229--256.

\bibitem{wu2017decentralized}
{\sc T.~Wu, K.~Yuan, Q.~Ling, W.~Yin, and A.~H. Sayed}, {\em Decentralized
  consensus optimization with asynchrony and delays}, IEEE TSIPN,  (2017),
  pp.~293--307.

\bibitem{wu2020safa}
{\sc W.~Wu, L.~He, W.~Lin, R.~Mao, C.~Maple, and S.~Jarvis}, {\em {SAFA}: A
  semi-asynchronous protocol for fast federated learning with low overhead},
  TC,  (2020), pp.~655--668.

\bibitem{xie2019asynchronous}
{\sc C.~Xie, S.~Koyejo, and I.~Gupta}, {\em Asynchronous federated
  optimization}, arXiv preprint, abs/1903.03934 (2019).

\bibitem{yang2018applied}
{\sc T.~Yang, G.~Andrew, H.~Eichner, H.~Sun, W.~Li, N.~Kong, D.~Ramage, and
  F.~Beaufays}, {\em Applied federated learning: Improving google keyboard
  query suggestions}, arXiv:1812.02903,  (2018).

\bibitem{yao2020pyhessian}
{\sc Z.~Yao, A.~Gholami, K.~Keutzer, and M.~W. Mahoney}, {\em Pyhessian: Neural
  networks through the lens of the hessian}, in Big Data, 2020, pp.~581--590.

\bibitem{ye2022decentralized}
{\sc H.~Ye, L.~Liang, and G.~Y. Li}, {\em Decentralized federated learning with
  unreliable communications}, JSTSP,  (2022), pp.~487--500.

\bibitem{ying2021exponential}
{\sc B.~Ying, K.~Yuan, Y.~Chen, H.~Hu, P.~Pan, and W.~Yin}, {\em Exponential
  graph is provably efficient for decentralized deep training}, NeurIPS,
  (2021), pp.~13975--13987.

\bibitem{yu2022hessian}
{\sc S.~Yu, Z.~Yao, A.~Gholami, Z.~Dong, S.~Kim, M.~W. Mahoney, and
  K.~Keutzer}, {\em Hessian-aware pruning and optimal neural implant}, in CVPR,
  2022, pp.~3880--3891.

\bibitem{yuan2016convergence}
{\sc K.~Yuan, Q.~Ling, and W.~Yin}, {\em On the convergence of decentralized
  gradient descent}, SIAM Journal on Optimization, 26 (2016), pp.~1835--1854.

\bibitem{Zhang2022FedDUAP}
{\sc H.~Zhang, J.~Liu, J.~Jia, Y.~Zhou, H.~Dai, and D.~Dou}, {\em Fedduap:
  Federated learning with dynamic update and adaptive pruning using shared data
  on the server}, in IJCAI, 2022, pp.~2776--2782.

\bibitem{zhang2021unified}
{\sc T.~Zhang, X.~Ma, Z.~Zhan, S.~Zhou, C.~Ding, M.~Fardad, and Y.~Wang}, {\em
  A unified dnn weight pruning framework using reweighted optimization
  methods}, in DAC, ACM, 2021, pp.~493--498.

\bibitem{WeiStaleness}
{\sc W.~Zhang, S.~Gupta, X.~Lian, and J.~Liu}, {\em Staleness-aware async-sgd
  for distributed deep learning}, in IJCAI, 2016, p.~2350–2356.

\bibitem{zhang2021validating}
{\sc Z.~Zhang, J.~Jin, Z.~Zhang, Y.~Zhou, X.~Zhao, J.~Ren, J.~Liu, L.~Wu,
  R.~Jin, and D.~Dou}, {\em Validating the lottery ticket hypothesis with
  inertial manifold theory}, NeurIPS, 34 (2021), pp.~30196--30210.

\bibitem{liu2022Efficient}
{\sc C.~Zhou, J.~Liu, J.~Jia, J.~Zhou, Y.~Zhou, H.~Dai, and D.~Dou}, {\em
  Efficient device scheduling with multi-job federated learning}, AAAI, 36
  (2022), pp.~9971--9979.

\bibitem{zhou2021distilled}
{\sc Y.~Zhou, G.~Pu, X.~Ma, X.~Li, and D.~Wu}, {\em Distilled one-shot
  federated learning}, arXiv:2009.07999,  (2021), pp.~1--16.

\bibitem{zhu2022topology}
{\sc T.~Zhu, F.~He, L.~Zhang, Z.~Niu, M.~Song, and D.~Tao}, {\em Topology-aware
  generalization of decentralized sgd}, in ICML, 2022, pp.~27479--27503.

\bibitem{zinkevich2010parallelized}
{\sc M.~Zinkevich, M.~Weimer, A.~J. Smola, and L.~Li}, {\em Parallelized
  stochastic gradient descent.}, in NeurIPS, vol.~23, 2010, pp.~1--37.

\bibitem{Zoph2017Neural}
{\sc B.~Zoph and Q.~V. Le}, {\em Neural architecture search with reinforcement
  learning}, in ICLR, 2017, pp.~1--16.

\end{thebibliography}
}

\newpage
\small
\appendix
\counterwithin{figure}{section}
\counterwithin{table}{section}

\section{Appendix}

In this section, we first present the experimental setup. Then, we show the details for the calculation of partial derivatives for dynamic weight update. Afterward, we explain the heterogeneous model aggregation. Furthermore, we analyze the impact of the current loss value brought by the pruning operation. Finally, we demonstrate the extra experimental results.

\subsection{Calculation of Partial Derivatives}

\begin{equation}
\begin{aligned}
&\nabla_{\lambda^{t_i-1}_{i,j}} F_i(m^{t_i}_i) \\
= &~(\frac{\partial F_i(m^{t_i}_i)}{\partial m^{t_i}_i})^\mathrm{T} \frac{\partial m^{t_i}_i}{\partial \lambda^{t_i-1}_{i,j}} \\
= &~g_i^\mathrm{T} \frac{\partial m^{t_i}_i}{\partial \lambda^{t_i-1}_{i,j}} \\
= &~g_i^\mathrm{T} \frac{\partial \sum_{k = i\text{ or }k \in \mathcal{M}} \omega^{t_i - 1}_{i,k} m^{t_i - 1}_k}{\partial \lambda^{t_i-1}_{i,j}} \\
= &~g_i^\mathrm{T} \frac{\partial \omega^{t_i - 1}_{i,j} }{\partial \lambda^{t_i-1}_{i,j}}m^{t_i - 1}_j \\
= &~g_i^\mathrm{T} \frac{\partial \frac{\omega^{'t_i-1}_{i,j}}{\sum_{k = i\text{ or }k \in \mathcal{M}} \omega^{'t_i-1}_{i,k}} }{\partial \lambda^{t_i-1}_{i,j}}m^{t_i - 1}_j \\
= &~g_i^\mathrm{T} \frac{\partial \frac{\omega^{'t_i-1}_{i,j}}{\sum_{k = i\text{ or }k \in \mathcal{M}} \omega^{'t_i-1}_{i,k}} }{\partial \omega^{'t_i-1}_{i,j}}\frac{\partial \omega^{'t_i-1}_{i,j}}{\partial \lambda^{t_i-1}_{i,j}}m^{t_i - 1}_j \\
= &~g_i^\mathrm{T} \frac{\partial (1 - \frac{\sum_{k = i\text{ or }(k \in \mathcal{M}\text{ \& }k \neq j)} \omega^{'t_i-1}_{i,k}}{\sum_{k = i\text{ or }k \in \mathcal{M}} \omega^{'t_i-1}_{i,k}} )}{\partial \omega^{'t_i-1}_{i,j}}\frac{\partial \omega^{'t_i-1}_{i,j}}{\partial \lambda^{t_i-1}_{i,j}}m^{t_i - 1}_j \\
= &~g_i^\mathrm{T} \frac{\sum_{k = i\text{ or }(k \in \mathcal{M}\text{ \& }k \neq j)} \omega^{'t_i-1}_{i,k}}{(\sum_{k = i\text{ or }k \in \mathcal{M}} \omega^{'t_i-1}_{i,k})^2} \frac{\partial \omega^{'t_i-1}_{i,j}}{\partial \lambda^{t_i-1}_{i,j}}m^{t_i - 1}_j \\
= &~g_i^\mathrm{T} \frac{\sum_{k = i\text{ or }(k \in \mathcal{M}\text{ \& }k \neq j)} \omega^{'t_i-1}_{i,k}}{(\sum_{k = i\text{ or }k \in \mathcal{M}} \omega^{'t_i-1}_{i,k})^2} \frac{\partial \frac{s_j * \lambda^{t_i-1}_{i,j}}{\sqrt{\bigtriangleup t^{t_i-1}_{i,j}} * loss^{t_i-1}_j}}{\partial \lambda^{t_i-1}_{i,j}}m^{t_i - 1}_j \\
= &~\frac{\sum_{k = i\text{ or }(k \in \mathcal{M}\text{ \& }k \neq j)} \omega^{'t_i-1}_{i,k}}{(\sum_{k = i\text{ or }k \in \mathcal{M}} \omega^{'t_i-1}_{i,k})^2} \frac{s_j g_i^\mathrm{T} m^{t_i - 1}_j}{\sqrt{\bigtriangleup t^{t_i-1}_{i,j}} * loss^{t_i-1}_j} \\
\end{aligned}
\end{equation}


\begin{figure}[!t]
\centering
\includegraphics[width=\linewidth]{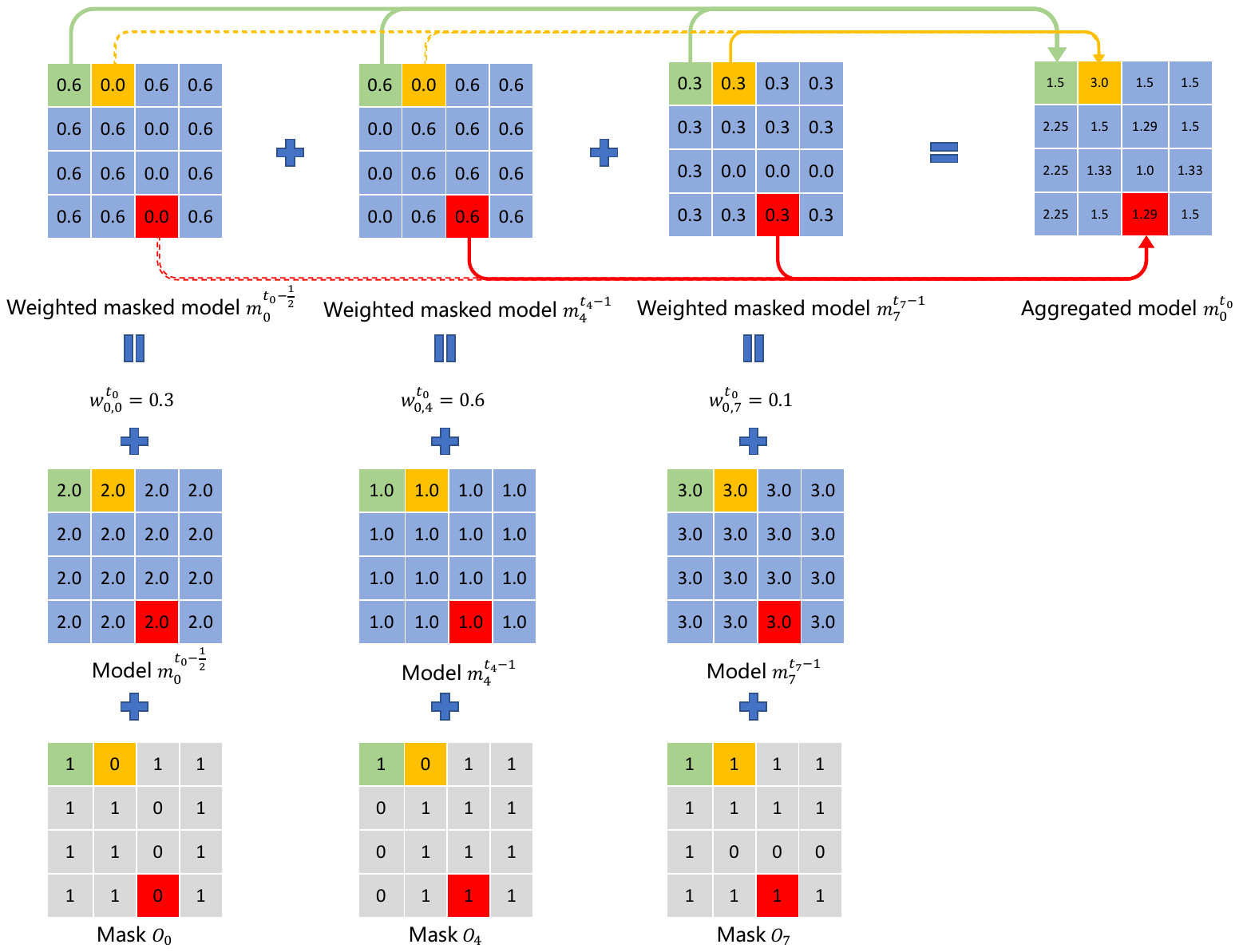}
\vspace{-2mm}
\caption{The heterogeneous model aggregation process. The dashed line represents that the parameter does not contributed to the aggregated model.}
\label{fig:aggregation}
\vspace{-5mm}
\end{figure}

\subsection{Heterogeneous Model Aggregation}

The dynamic model aggregation contains three stages. 
First, we exploit a reinforcement learning-based model to select proper neighbor models for the aggregation. Then, we dynamically update the weights of the selected neighbor models. Finally, we merge the heterogeneous models with the updated weights to a new model.

As shown in Figure \ref{fig:aggregation}, we consider a system of 8 devices. There are three neighbor models, i.e., Models $m_4$, $m_6$, $m_7$ for Devices $4$, $6$, $7$, respectively, on Device $0$. After the local training process, we have $m_0^{t_0 - \frac{1}{2}}$. Then, after the model selection process, Models $m_4$ and $m_7$ are kept for aggregation. In addition, we have the updated weights $\omega^{t_0}_{0,0} = 0.3$, $\omega^{t_0}_{0,4} = 0.6$, $\omega^{t_0}_{0,7} = 0.1$, and masks $O_0$, $O_4$, $O_7$. Afterward, the parameter in the first row and the first column is calculated: (2 * 0.3 * 1 + 1 * 0.6 * 1 + 3 * 0.1 * 1)(0.3 + 0.6 + 0.1) = 1.5 (based on Formula \ref{eq:parameterAggregation}), and the same for other parameters.

\subsection{The Impact of the Current Loss Value}

We can have the following formula by exploiting the Lagrangian method to find a saddle point:
\begin{align}
\label{eq:lagrangeDef}
    \mathcal{L} = &~g^T_i \Delta m_i + \frac{1}{2} \Delta m^{T}_i \mathcal{H} \Delta m_i + \lambda^T(\Delta m_i^p + m_i^p), \nonumber\\
    &~\frac{\partial \mathcal{L}}{\partial \Delta m_i} = g^T_i + \mathcal{H} \Delta m_i + \begin{pmatrix} \lambda \\ 0 \end{pmatrix} = 0,
\end{align}
where $\lambda \in \mathbb R^p$ is the Lagrange multiplier. Then, we express the pruning and the remained parts with $p$ and $r$ and get Formula \ref{eq:lagrangeExp}.
\begin{align}
\label{eq:lagrangeExp}
    \begin{pmatrix} g_i^p \\ g_i^r \end{pmatrix}^T + \begin{pmatrix} \mathcal{H}^{p,p},\mathcal{H}^{p,r} \\  \mathcal{H}^{r,p},\mathcal{H}^{r,r} \end{pmatrix} \begin{pmatrix}\Delta m_i^p \\ \Delta m_i^r \end{pmatrix} + \begin{pmatrix} \lambda \\ 0 \end{pmatrix} = 0.
\end{align}
In addition, we have $\Delta m_i^p = - m_i^p$ and plug this into Formula \ref{eq:lagrangeExp} to get Formula \ref{eq:lagrangePlug}.
\begin{align}
\label{eq:lagrangePlug}
    g_i^r - \mathcal{H}^{r,p} m_i^p +\mathcal{H}^{r,r} \Delta m_i^r = 0.
\end{align}
And, we can get the expression of $\Delta m_i^r$ as shown in Formula \ref{eq:mi}.
\begin{align}
\label{eq:mi}
    \Delta m_i^r = (\mathcal{H}^{r,r})^{-1} \mathcal{H}^{r,p} m_i^p - (\mathcal{H}^{r,r})^{-1} g_i^r.
\end{align}

\begin{figure*}[!t]
\centering
\subfigure[Emnist \& LeNet]{
\includegraphics[width=0.31\textwidth]{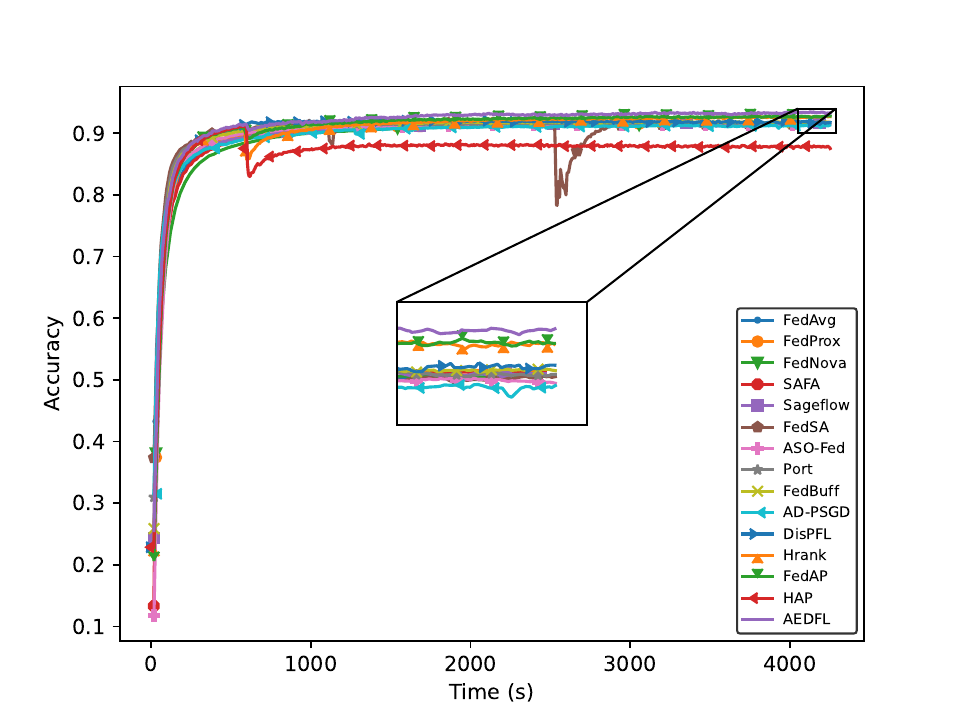}
\label{sub:emnist_LeNet}
}
\subfigure[CIFAR \& ResNet]{
\includegraphics[width=0.31\textwidth]{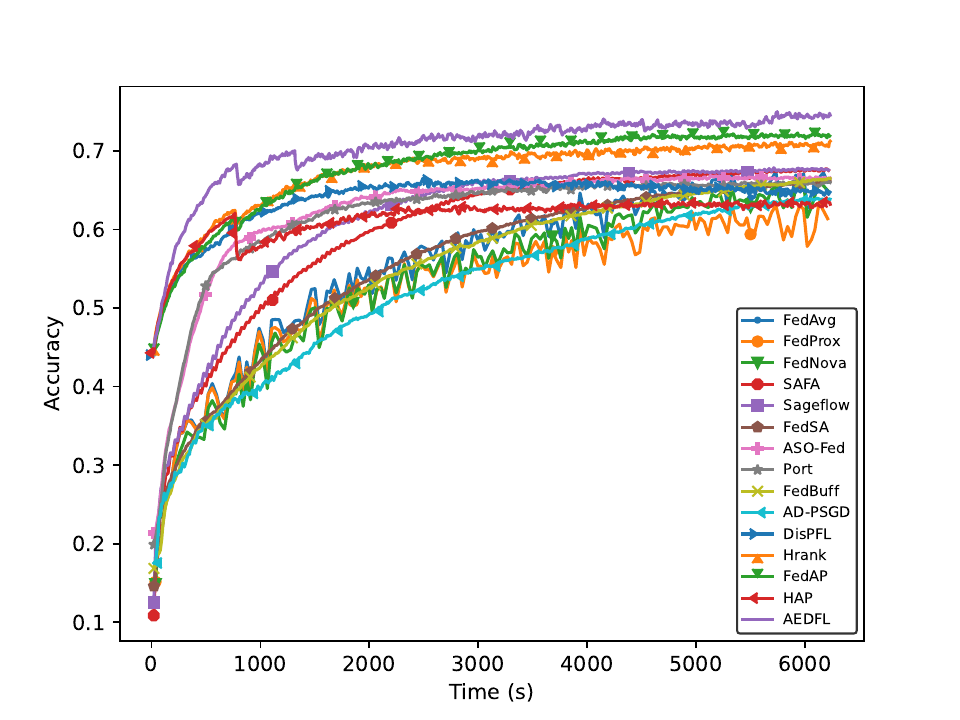}
\label{sub:cifar_resnet}
}
\subfigure[Tiny-ImageNet \& VGG]{
\includegraphics[width=0.31\textwidth]{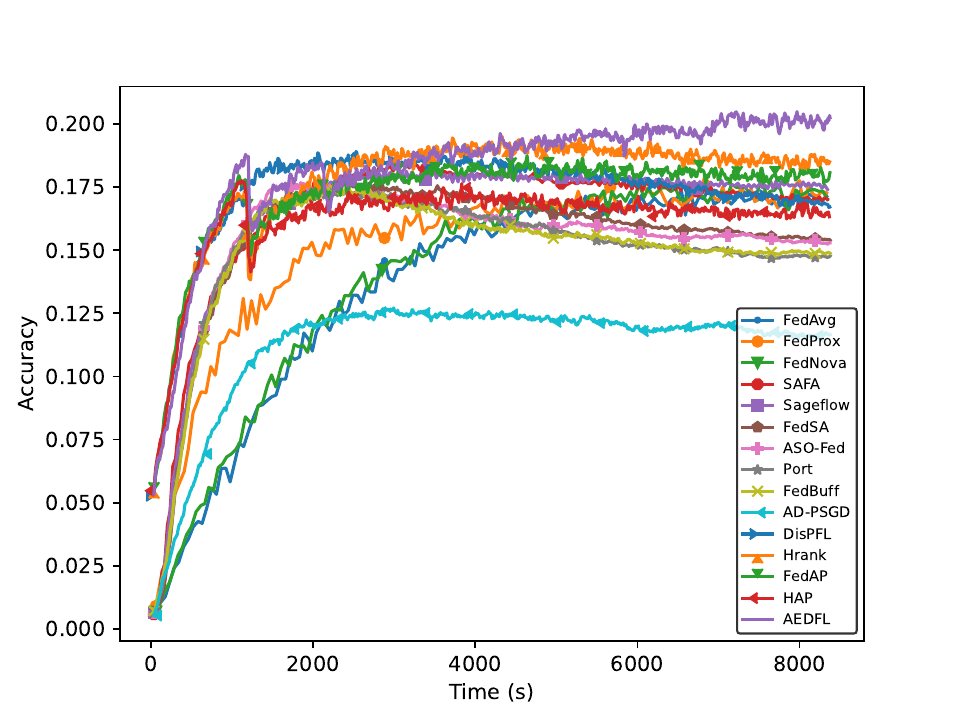}
\label{sub:tinyImageNet_VGG}
} 
\vspace*{-3mm}
\caption{The accuracy and training time with \TheName{} and diverse baseline approaches.}
\vspace*{-2mm}
\label{fig:accuracyTime}
\end{figure*}

Finally, we can calculate the impact on the current loss value by Formula \ref{eq:finalLoss}. Please note that a Hessian matrix is symmetric, i.e. $((\mathcal{H}^{r,r})^{-1})^T = (\mathcal{H}^{r,r})^{-1}$ and $(\mathcal{H}^{r,p})^T = \mathcal{H}^{p,r}$.
\begin{align}
\label{eq:finalLoss}
    \Delta F_i^C = &~g^T_i \Delta m_i + \frac{1}{2} \Delta m^{T}_i \mathcal{H} \Delta m_i \nonumber\\
    = &~\frac{1}{2} \begin{pmatrix}\Delta m_i^p \\ \Delta m_i^r \end{pmatrix}^T \begin{pmatrix} \mathcal{H}^{p,p},\mathcal{H}^{p,r} \\  \mathcal{H}^{r,p},\mathcal{H}^{r,r} \end{pmatrix} \begin{pmatrix}\Delta m_i^p \\ \Delta m_i^r \end{pmatrix} \nonumber\\
    &~+ \begin{pmatrix} g_i^p \\ g_i^r \end{pmatrix}^T \begin{pmatrix} \Delta m_i^p \\ \Delta m_i^r \end{pmatrix}^T \nonumber\\
    = &~\frac{1}{2} ( (\Delta m_i^p)^T \mathcal{H}^{p,p} \Delta m_i^p + (\Delta m_i^r)^T \mathcal{H}^{r,p} \Delta m_i^p ) \nonumber\\
    &~+ \frac{1}{2}( (\Delta m_i^p)^T \mathcal{H}^{p,r} \Delta m_i^r + (\Delta m_i^r)^T \mathcal{H}^{r,r} \Delta m_i^r ) \nonumber\\
    &~ + (g_i^p)^T \Delta m_i^p + (g_i^r)^T \Delta m_i^r \nonumber \\
    = &~\frac{1}{2} (-m_i^p)^T \mathcal{H}^{p,p} (-m_i^p) \nonumber\\
    &~+ \frac{1}{2}((\mathcal{H}^{r,r})^{-1} \mathcal{H}^{r,p} m_i^p - (\mathcal{H}^{r,r})^{-1} g_i^r)^T \mathcal{H}^{r,p} (-m_i^p) \nonumber\\
    &~+ \frac{1}{2}(-m_i^p)^T \mathcal{H}^{p,r} ((\mathcal{H}^{r,r})^{-1} \mathcal{H}^{r,p} m_i^p - (\mathcal{H}^{r,r})^{-1} g_i^r) \nonumber\\
    &~+ \frac{1}{2}(((\mathcal{H}^{r,r})^{-1} \mathcal{H}^{r,p} m_i^p - (\mathcal{H}^{r,r})^{-1} g_i^r)^T \mathcal{H}^{r,r}\nonumber\\
    &\quad\quad((\mathcal{H}^{r,r})^{-1} \mathcal{H}^{r,p} m_i^p - (\mathcal{H}^{r,r})^{-1} g_i^r) ) \nonumber\\
    &~+ (g_i^r)^T ((\mathcal{H}^{r,r})^{-1} \mathcal{H}^{r,p} m_i^p - (\mathcal{H}^{r,r})^{-1} g_i^r) \nonumber \\
    &~+ (g_i^p)^T (-m_i^p) \nonumber \\
    = &~\frac{1}{2} (m_i^p)^T \mathcal{H}^{p,p} m_i^p \nonumber\\
    &~- \frac{1}{2}(m_i^p)^T \mathcal{H}^{p,r} (\mathcal{H}^{r,r})^{-1} \mathcal{H}^{r,p} m_i^p \nonumber\\
    &~- \frac{1}{2}(g_i^r)^T (\mathcal{H}^{r,r})^{-1} g_i^r \nonumber\\
    &~+ (g_i^r)^T (\mathcal{H}^{r,r})^{-1} \mathcal{H}^{r,p} m_i^p \nonumber \\
    &~- (g_i^p)^T m_i^p \nonumber \\
\end{align}


\begin{figure}[!t]
\centering
\includegraphics[width=\linewidth]{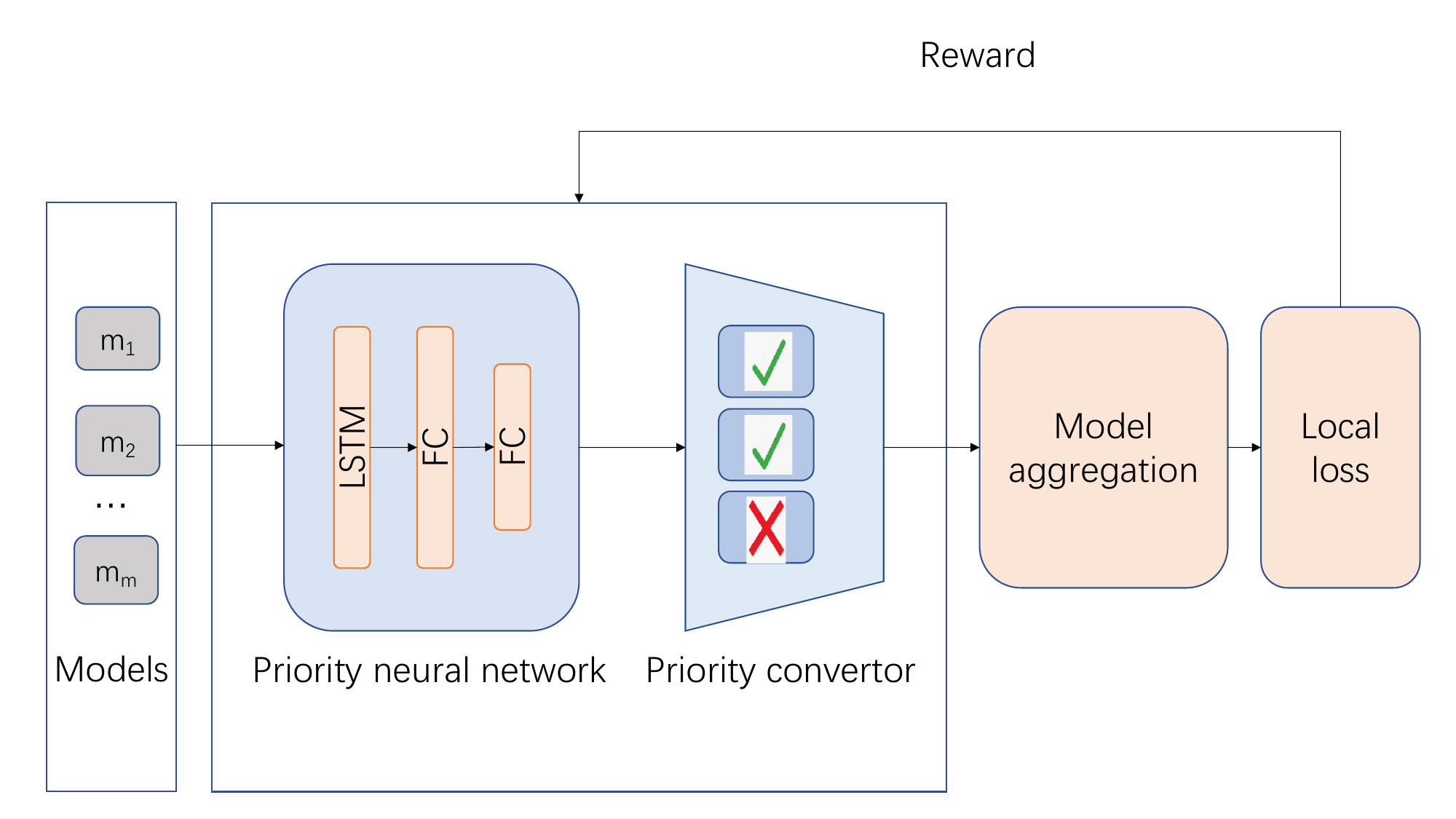}
\vspace{-6mm}
\caption{The architecture of the reinforcement learning model.}
\vspace{-1mm}
\label{fig:RL}
\end{figure}

\begin{figure*}[!t]
\centering
\subfigure[CIFAR \& LeNet]{
\includegraphics[width=0.31\textwidth]{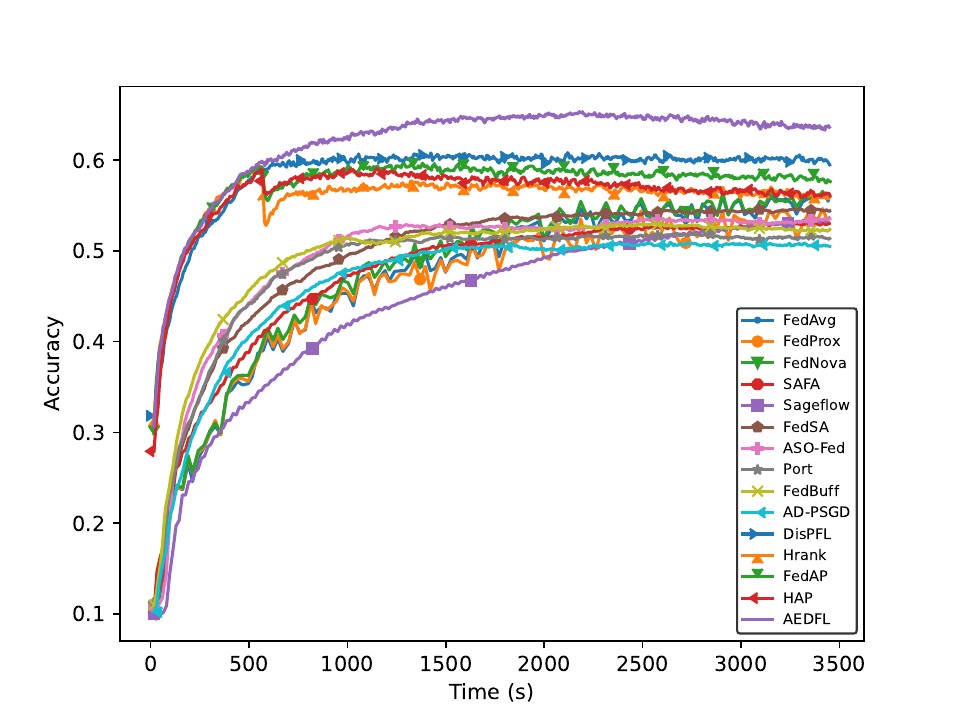}
\label{sub:cifar_LeNet}
}
\subfigure[CIFAR \& VGG]{
\includegraphics[width=0.31\textwidth]{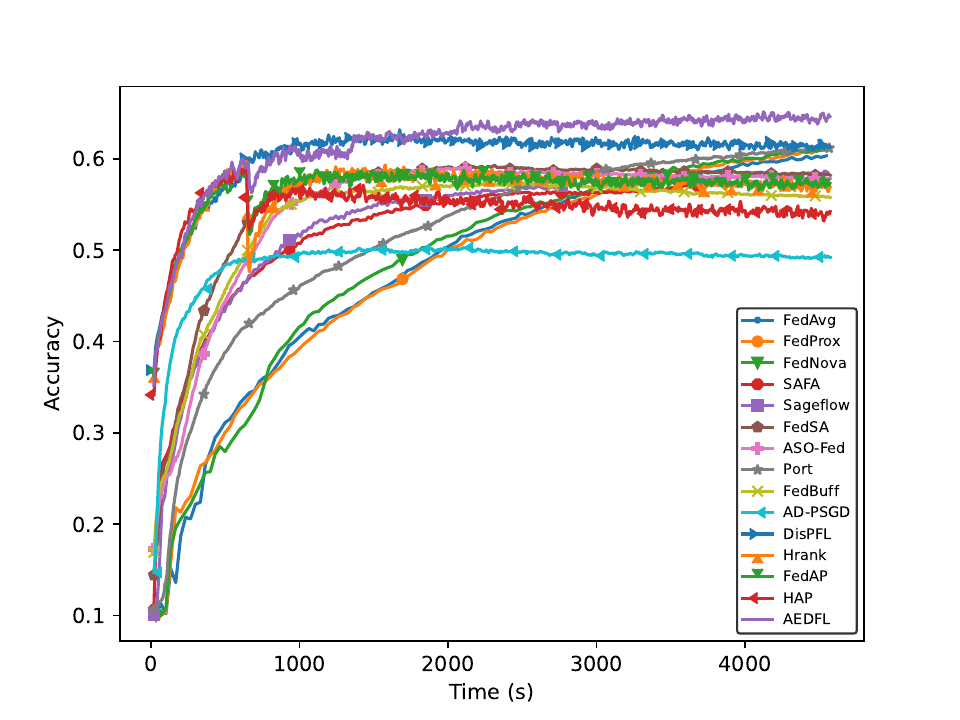}
\label{sub:cifar_vgg}
}
\subfigure[Tiny-ImageNet \& ResNet]{
\includegraphics[width=0.31\textwidth]{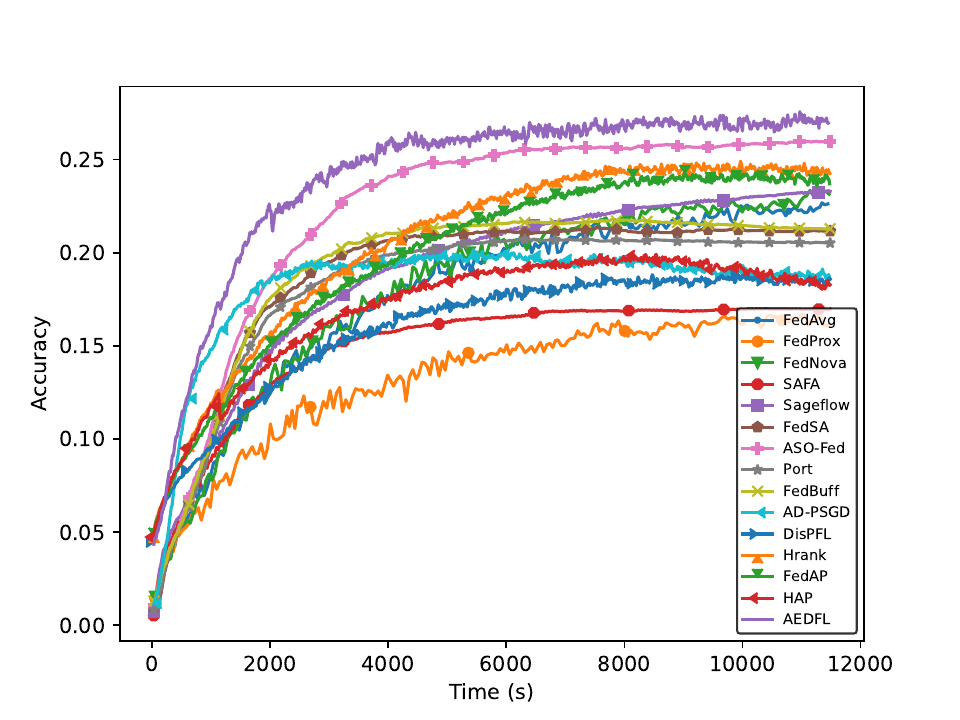}
\label{sub:tinyImageNet_ResNet}
}
\caption{The accuracy and training time with \TheName{} and diverse baseline approaches.}
\label{fig:accuracyTimeA}
\end{figure*}

\subsection{Adaptive Pruning}
\label{sec:adaptive_pruning}

\begin{figure}[!t]
\centering
\includegraphics[width=0.45\textwidth]{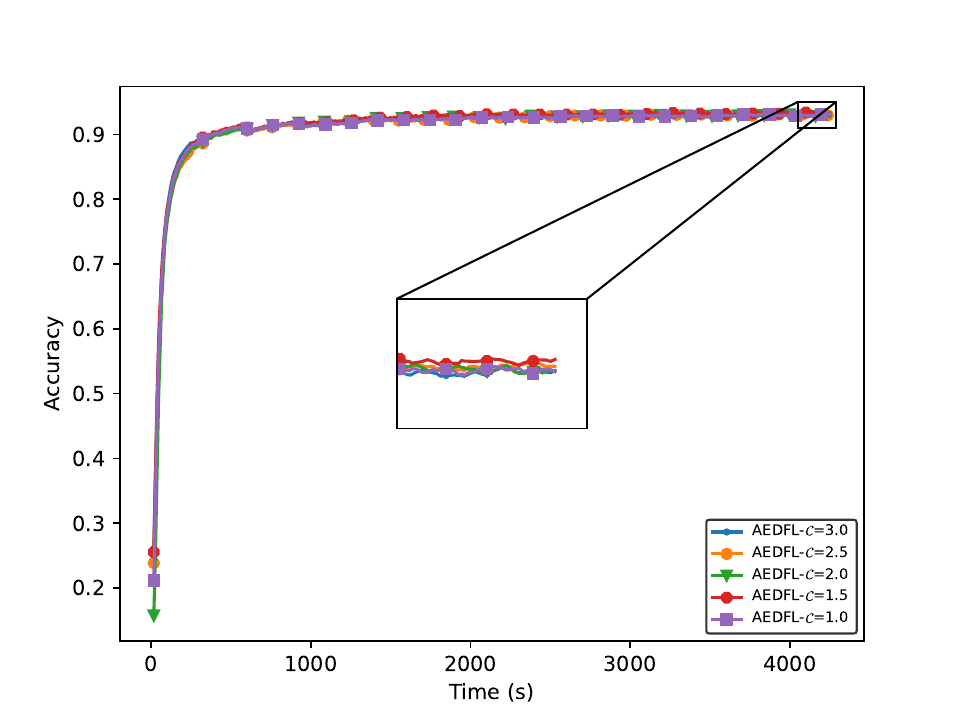}
\caption{The impact of hyper-parameter $\mathcal{C}$ with EMNIST \& LeNet.}
\label{fig:cimpact_emnist_lenet}
\end{figure}

\begin{figure}[!t]
\centering
\includegraphics[width=0.45\textwidth]{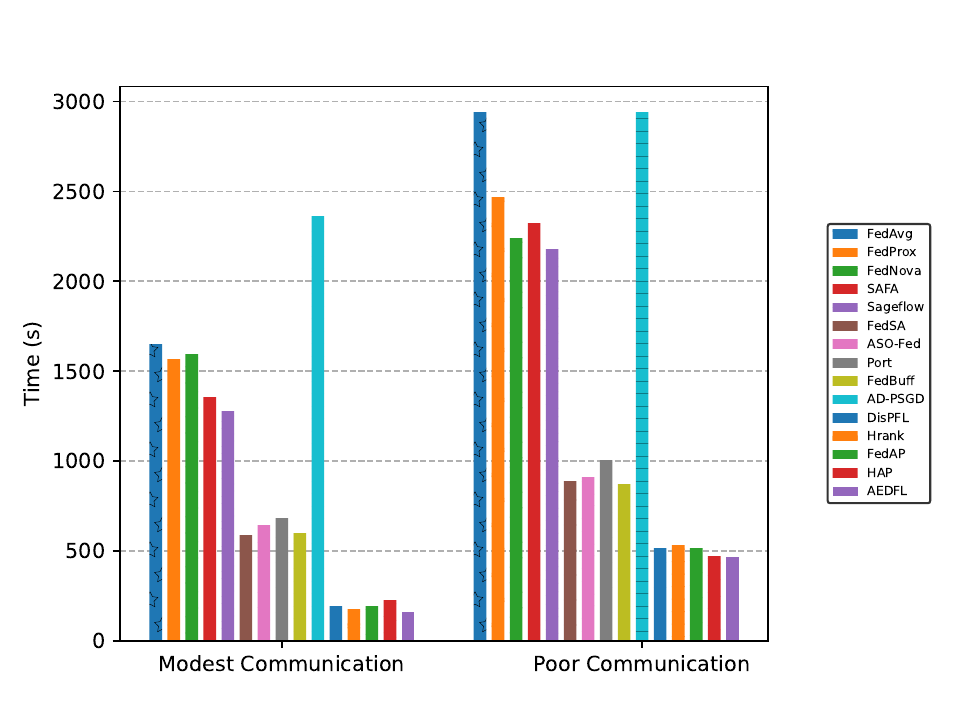}
\caption{Time to target accuracy (0.5) under diverse bandwidth with CIFAR \& LeNet.}
\label{fig:modestNetwork}
\end{figure}

\begin{figure}[h]
\vspace{-4mm}
\begin{algorithm}[H]
\caption{Adaptive Pruning}
\label{alg:pruning}
\begin{algorithmic}[1]
\REQUIRE  \quad \newline
$i$: The index of the device \newline
$m^o_i$: The initialized model on Device $i$ \newline
$m_i$: The current model on Device $i$ \newline
$M$: The set of neighbor models \newline
$P$: The set of pruning rates of neighbor models \newline
$K$: The number of parameters in $m_i$
\ENSURE \quad \newline
$m^p_i$: The pruned model
\STATE Calculate $p_i$ with $P$ based on Formula \ref{eq:pruningRateAggregation}
\STATE Sort parameters in $m_i$ to $[\mu^1_i, \mu^2_i,\cdots,\mu^K_i]$ in ascending order according to Formula \ref{eq:deltaLossTotal}
\STATE $m^p_i \gets$ Prune the first $K*p_i$ parameters 
\end{algorithmic}
\end{algorithm}
\vspace{-8mm}
\end{figure}

As shown in Algorithm \ref{alg:pruning}, we first calculate the pruning rate of the full model (Line 1). Then, we sort the parameters in ascending order based on Formula \ref{eq:deltaLossTotal} (Line 2). Afterwards, we prune the first $K*p_i$ parameters with $K$ representing the number of parameters (Line 3). This algorithm is carried out every certain local updates to dynamically adjust the model.

\subsection{Experimental Setup}

In the experiment, we set the learning rate to 0.03 for the local update, the learning rate decay to 0.001, the batch size to 50, and the local epoch to 4. We utilize the Dirichlet distribution \cite{li2021federated} (with 0.5 as the concentration parameter) to partition the data and attribute a certain number of samples into each device according to a lognormal distribution (with $\frac{s}{n}$ as the mean and 0.1 as the standard deviation, where $s$ is the total number of samples among all Devices and $n$ is the total number of Devices). The distribution of users' staleness is uniform, and we have set a disparity of 15 times between the maximum and minimum staleness. All the components (including the model update, the inference, and the update of the reinforcement learning model for model selection, dynamic weight update, and model pruning for sparse training) are considered for calculating the training time and communication costs. The major notations 
and the values for other hyper-parameters 
are not shown due to page limit, but can be provided upon request (see details in \url{https://anonymous.4open.science/r/AEDFL-8FD5/}).

FedAvg, FedProx, and FedNova are state-of-the-art centralized synchronous FL approaches. SAFA and Sageflow are state-of-the-art asynchronous approaches, and we adapt these approaches in \TheName{}. FedSA, ASO-Fed, Port, and FedBuff are state-of-the-art asynchronous distributed FL approaches. AD-PSGD and DisPFL are state-of-the-art decentralized FL approaches. Hrank, FedAP, and HAP are pruning approaches, which are adapted to our framework. DisPFL exploits a pruning method for an efficient training process as well.

\subsection{Extra Experimental Results}

\begin{figure}[!t]
\centering
\includegraphics[width=0.45\textwidth]{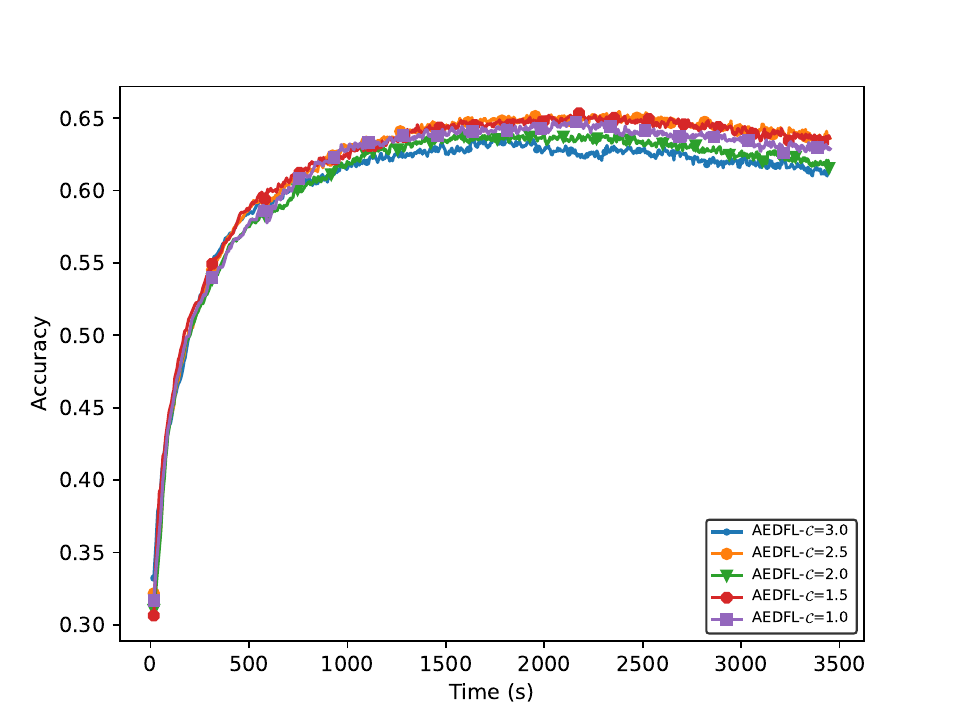}
\caption{The impact of hyper-parameter $\mathcal{C}$ with CIFAR \& LeNet.}
\label{fig:cimpact_cifar_lenet}
\end{figure}

\begin{figure}[!t]
\centering
\includegraphics[width=0.45\textwidth]{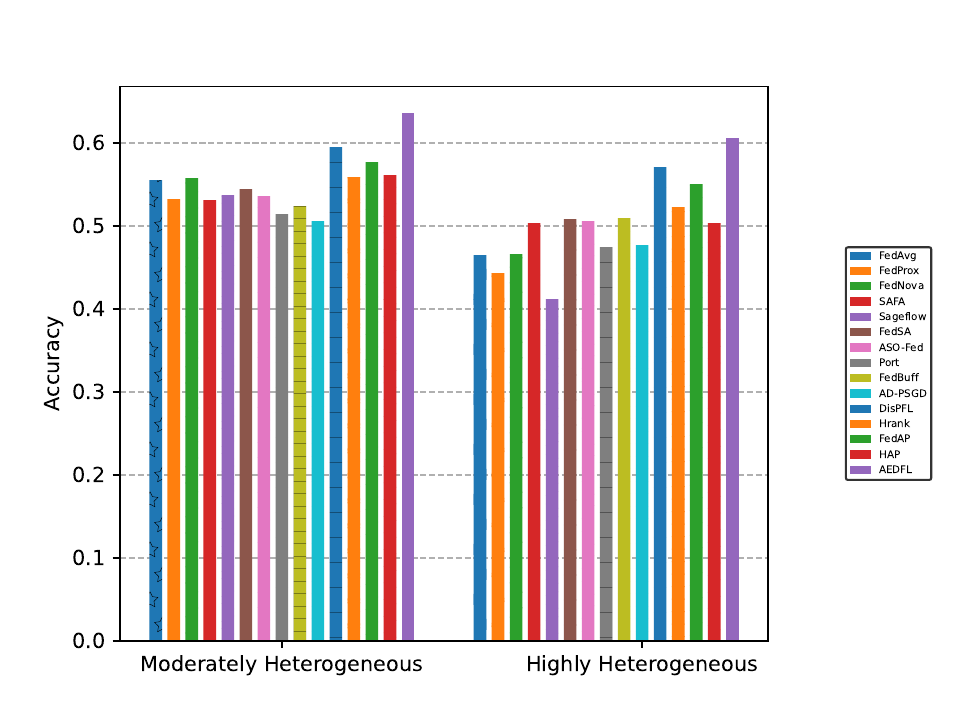}
\caption{The accuracy of \TheName{} and diverse baseline approaches under various device heterogeneity with CIFAR \& LeNet.}
\vspace*{-4mm}
\label{fig:heterogeneity}
\end{figure}

The accuracy, training time, and computation costs of \TheName{} and diverse baseline approaches corresponding to CIFAR with LeNet and VGG are shown in Table \ref{tab:cmp_ASDFL_A}. The advantages of \TheName{} can be up to 8.1\% for FedAvg, 10.5\% for FedProx, 7.9\% for FedNova,  10.6\% for SAFA, 10\% for Sageflow, 9.2\% for FedSA, 10.1\% for ASO-Fed, 12.3\% for Port, 11.3\% for FedBuff, 13.1\% for AD-PSGD, 4.1\% for DisPFL, 7.8\% for Hrank 6\% for FedAP, and 7.6\% for HAP in terms of accuracy. Furthermore, \TheName{} corresponds to a short training time. The training time of \TheName{} to achieve a target accuracy can be up to 89.6\% (compared with FedAvg), 89.6\% (compared with FedProx), 87.3\% (compared with FedNova), 87.6\% (compared with SAFA), 92\% (compared with Sageflow), 84\% (compared with FedSA), 80.4\% (compared with ASO-Fed), 80.8\% (compared with Port), 78.6\% (compared with FedBuff), 88.5\% (compared with AD-PSGD), 19.5\% (compared with DisPFL), 7.3\% (compared with Hrank), 1.3\% (compared with FedAP), and 10.1\% (compared with HAP) shorter. In addition, \TheName{} corresponds to much smaller computation costs (up to 33\% compared with FedAvg, FedProx, FedNova, SAFA, Sageflow, FedSA, ASO-Fed, Port, FedBuff, AD-PSGD, 16.3\% compared with DisPFL, 31.9\% compared with Hrank, 31.8\% compared with FedAP, and 16.4\% compared with HAP) without accuracy degradation.

The accuracy and the training time with \TheName{} and diverse baseline approaches are shown in Figures \ref{fig:accuracyTime} and \ref{fig:accuracyTimeA}. From the figures, we can see \TheName{} corresponds to the fastest training speed and the accuracy is the highest among all the methods. The minimum advantages of \TheName{} are significant as well, i.e., 3.4\% higher, 42.3\% faster, and 7.2\% smaller, compared with diverse baseline approaches. Please note that although the convergence accuracy of some cases (e.g., HAP with Emnist and LeNet) in Table \ref{tab:cmp_ASDFL} is smaller than target accuracy, the model may achieve the target accuracy during the training and there is the time to achieve target accuracy. The accuracy may decrease during the training as shown in Figures \ref{fig:accuracyTime} and \ref{fig:accuracyTimeA}, which is aligned with recent observations (double descent) \cite{nakkiran2021deep,d2020double,adlam2020understanding}.

\begin{table}[!h]
  \scriptsize
  \caption{The accuracy, training time, and computation costs with \TheName{} and diverse baseline approaches. ``Acc'' represents the convergence accuracy. ``Time'' refers to the training time to achieve a target accuracy (0.5 for LeNet with CIFAR, 0.58 for VGG with CIFAR). ``MFLOPs'' represents the computational costs. ``/'' represents that training does not achieve the target accuracy.
  }
  \label{tab:cmp_ASDFL_A}
  \centering
  \begin{tabular}{l|lll|lll}
    \toprule
    \multirow{2}{*}{Method} & \multicolumn{3}{c|}{CIFAR \& LeNet}  & \multicolumn{3}{c}{CIFAR \& VGG} \\
    \cmidrule(r){2-7}  & Acc       & Time   & MFLOPS  & Acc       & Time   & MFLOPS   \\
    \midrule
    \TheName{}          &  \textbf{0.6363}    & \textbf{1478}   & \textbf{0.436} & \textbf{0.6462} &  \textbf{456}    & \textbf{2.98}   \\
    FedAvg              &  0.5553    & 14178   &  0.652  & 0.6032    &  3584    & 5.16  \\
    FedProx             &  0.5317    & 14178   &  0.652  & 0.6098    &  3518    & 5.16   \\
    FedNova             &  0.5575    & 11654   &  0.652  & 0.6101    &  3252    & 5.16  \\
    SAFA                &  0.5302    & 11900   &  0.652  & 0.5732    &  /       & 5.16  \\
    Sageflow            &  0.5361    & 18480   &  0.652  & 0.5849    &  3585    & 5.16   \\
    FedSA               &  0.5440    & 9254    &  0.652  & 0.5818    &  1308    & 5.16  \\
    ASO-Fed             &  0.5353    & 7539    &  0.652  & 0.5787    &  1415    & 5.16  \\
    Port                &  0.5136    & 7688    &  0.652  & 0.6127    &  2816    & 5.16  \\
    FedBuff             &  0.5236    & 6890    &  0.652  & 0.5576    &  /       & 5.16  \\
    AD-PSGD             &  0.5050    & 12811   &  0.652  & 0.4919    &  /       & 5.16  \\
    DisPFL              &  0.5919    & 1836    &  0.521  & 0.6127    &  577     & 3.11  \\
    Hrank               &  0.5587    & 1594    &  0.641  & 0.5689    &  488     & 3.18  \\
    FedAP               &  0.5765    & 1496    &  0.64   & 0.5730    &  551     & 3.11 \\
    HAP                 &  0.5607    & 1645    &  0.522  & 0.5421    &  480     & 3.36  \\
    \bottomrule
  \end{tabular}
  \vspace{-2mm}
\end{table}

\subsection{Performance Evaluation within Divers Environments}

In this section, we evaluate 
\TheName{} with diverse topologies, network, hyperparameter ($\mathcal{C}$), device heterogeneity, numbers of devices, and ablation study.

\textbf{Impact of network:} While edge devices have limited network connection, we analyze the performance of \TheName{} with diverse bandwidth. As shown in Figure \ref{fig:modestNetwork}, when the bandwidth becomes modest, \TheName{} corresponds to significantly higher accuracy (up to 12.8\%) and training speed (up to 92\%), as well. When the network connection becomes worse, the advantages of \TheName{} become significant because of communication reduction brought by the sparse training with the adaptive pruning operation.

\textbf{Impact of $\mathcal{C}$:} While the hyper-parameter $\mathcal{C}$ corresponds to the importance of the gradients during the pruning operation, we carry out experimentation with diverse $\mathcal{C}$. As shown in Figures \ref{fig:cimpact_emnist_lenet} and \ref{fig:cimpact_cifar_lenet}, when $\mathcal{C}=1.5$, the performance of \TheName{} is the best for both EMNIST \& LeNet and CIFAR \& LeNet (up to 2.2\% better than other values), which demonstrates that the exploitation is of much importance for the training process. In addition, we fine-tune the values of $\mathcal{C}$ and take the proper values for other datasets and models.

\begin{figure}[!t]
\centering
\includegraphics[width=0.45\textwidth]{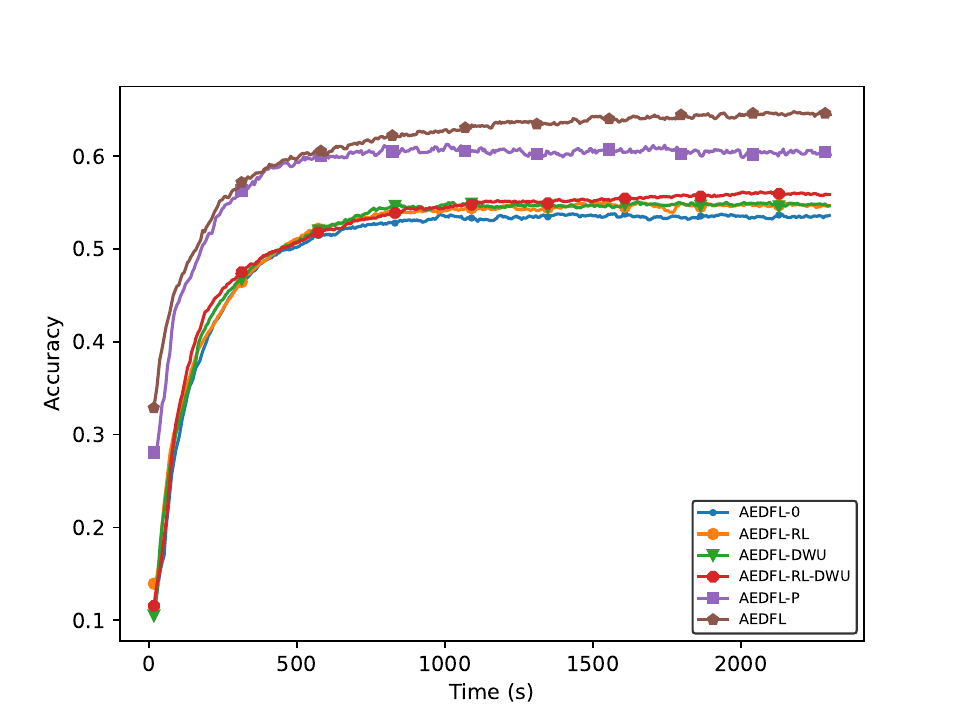}
\caption{The accuracy and training time of \TheName{} variants with CIFAR \& LeNet.}
\label{fig:ablation}
\end{figure}

\begin{figure}
\centering
\includegraphics[width=0.5\textwidth]{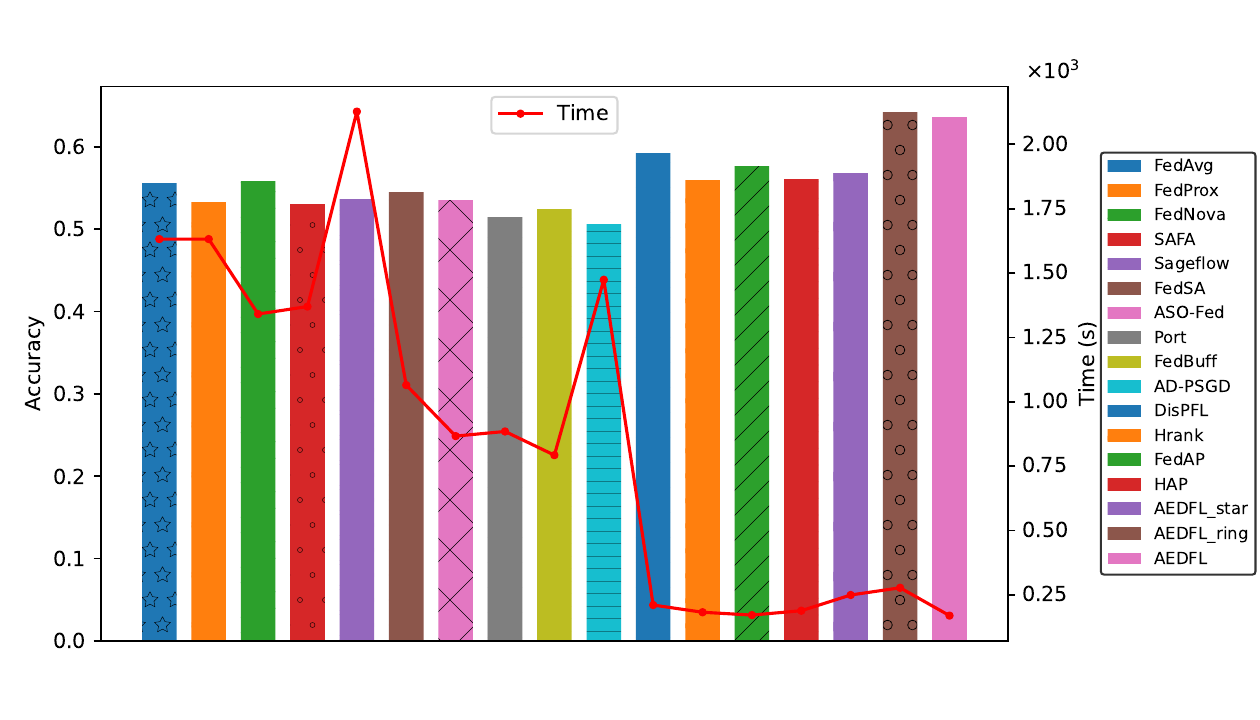}
\vspace{-3mm}
\caption{The accuracy and training time of \TheName{} and diverse baseline approaches with CIFAR \& LeNet and other topologies.}
\vspace{-6mm}
\label{fig:cifar10_lenet_topo}
\end{figure}

\textbf{Impact of device heterogeneity:} As shown in Figure \ref{fig:heterogeneity}, when the devices are heterogeneous (the standard deviation of the time to perform local update is significant), the advantage of \TheName{} becomes obvious (from 12.8\% to 19.6\%). When the devices are highly heterogeneous, \TheName{} significantly outperforms other baseline approaches in terms of accuracy because of dynamic weight update.

Furthermore, Figure \ref{fig:ablation} shows the accuracy and training time with diverse versions of \TheName{} for the ablation study. The accuracy of \TheName{}-RL is much higher (1.1\%) than \TheName{}-0 and \TheName{}-DWU can significantly outperform \TheName{}-0 in terms of accuracy (1\%). Additionally, the combination of the reinforcement learning-based model selection and the dynamic weight update corresponds to higher accuracy (2.3\%) than \TheName{}-0. Furthermore, \TheName{}-P can accelerate the training process (up to 61.2\% faster than \TheName{}-0 to achieve the target accuracy of 0.5) and lead to higher accuracy (6.5\% compared with \TheName{}-0).
Finally, \TheName{} achieves the highest accuracy (up to 10.9\% compared with \TheName{}-0) and efficiency (up to 64.9\% compared with \TheName{}-0 to achieve the target accuracy of 0.5 ).

\textbf{Impact of topologies:} We exploit the exponential topology because of its reported excellent performance \cite{assran2019stochastic} and it does not incur poor generalization capacity compared with other topologies \cite{zhu2022topology}. We carry out additional experiments with other topologies, i.e., Star and Ring, based on CIFAR and LeNet, and the results are shown in 
Figure \ref{fig:cifar10_lenet_topo}. Our approach with Ring still corresponds to excellent performance, e.g., up to 13.6\% higher accuracy and 88.2\% faster, compared with baseline approaches. Although Ring corresponds to higher accuracy compared with exponential, exponential can achieve the target accuracy within the shortest training time. In addition, the advantages of \TheName{} (with the exponential topology) can be up to 8.1\% for FedAvg, 10.5\% for FedProx, 7.9\% for FedNova,  10.6\% for SAFA, 10\% for Sageflow, 9.2\% for FedSA, 10.1\% for ASO-Fed, 12.3\% for Port, 11.3\% for FedBuff, 13.1\% for AD-PSGD, 4.1\% for DisPFL, 7.8\% for Hrank 6\% for FedAP, and 7.6\% for HAP in terms of accuracy. In addition, \TheName{} corresponds to a short training time. The training time of \TheName{} to achieve a target accuracy can be up to 89.6\% (compared with FedAvg), 89.6\% (compared with FedProx), 87.3\% (compared with FedNova), 87.6\% (compared with SAFA), 92\% (compared with Sageflow), 84\% (compared with FedSA), 80.4\% (compared with ASO-Fed), 80.8\% (compared with Port), 78.6\% (compared with FedBuff), 88.5\% (compared with AD-PSGD), 19.5\% (compared with DisPFL), 7.3\% (compared with Hrank), 1.3\% (compared with FedAP), and 10.1\% (compared with HAP) shorter. Furthermore, \TheName{} corresponds to much smaller computation costs (up to 33\% compared with FedAvg, FedProx, FedNova, SAFA, Sageflow, FedSA, ASO-Fed, Port, FedBuff, AD-PSGD, 16.3\% compared with DisPFL, 31.9\% compared with Hrank, 31.8\% compared with FedAP, and 16.4\% compared with HAP) without accuracy degradation.

\begin{figure}
\centering
\includegraphics[width=0.45\textwidth]{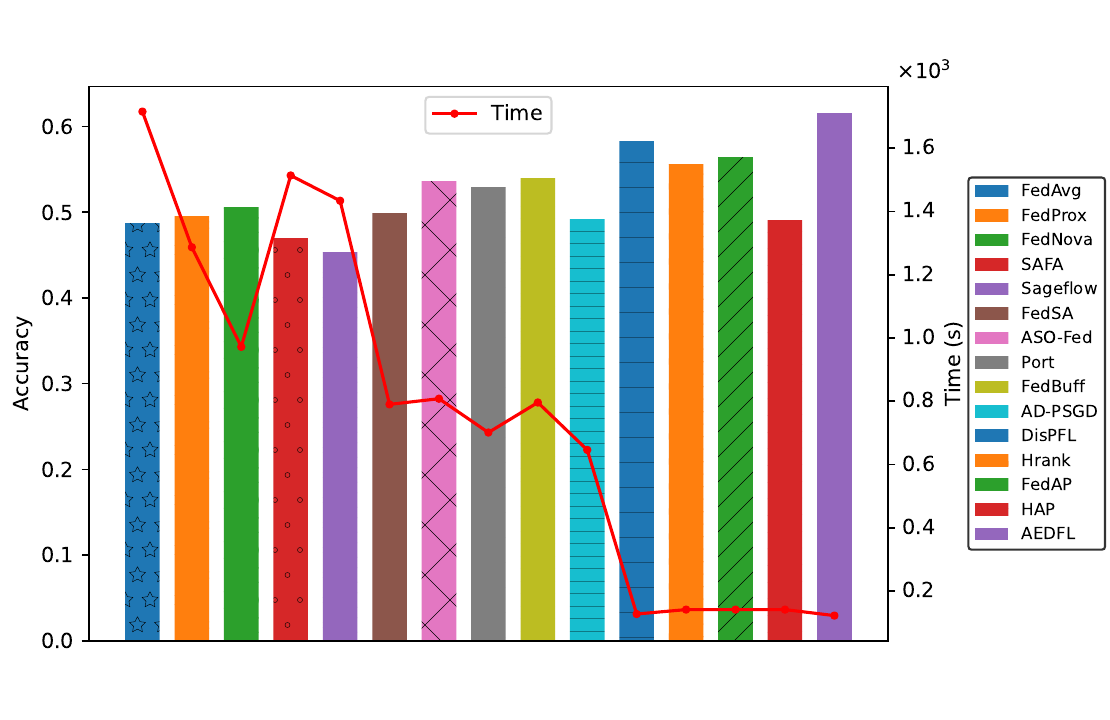}
\caption{The accuracy and training time of \TheName{} and diverse baseline approaches with CIFAR \& LeNet and 200 clients.}
\vspace{-5mm}
\label{fig:cifar10_200clients}
\end{figure}

\textbf{Impact of the number of devices:}
The advantages of \TheName{} become more significant when the number of devices becomes bigger. We change the number of devices $n$ to 200 in the CIFAR \& LeNet experiment and leave other hyper-parameters unchanged 
to evaluate the impact of the number of devices on \TheName{}. As shown in 
Figure \ref{fig:cifar10_200clients}, \TheName{} achieves up to 16.3\% higher accuracy and 92.9\% faster training time compared with other baselines. As we exploit the exponential topology and sample the neighbors to send updated models, the scalability of \TheName{} with different numbers of devices remains excellent.

\subsection{Ablation Study}

Figure \ref{fig:ablation} shows the accuracy and training time with diverse versions of \TheName{} for the ablation study. The accuracy of \TheName{}-RL is much higher (1.1\%) than \TheName{}-0, which demonstrates that the reinforcement learning-based model selection can significantly improve the accuracy. In addition, \TheName{}-DWU can significantly outperform \TheName{}-0 in terms of accuracy (1\%), which shows the advantage of the dynamic weight update. Additionally, the combination of the reinforcement learning-based model selection and the dynamic weight update corresponds to the highest accuracy (2.3\% higher than \TheName{}-0). This demonstrates the application of reinforcement learning-based model selection can automatically select proper models for aggregation so as to reduce the search space of the dynamic weight update while leading to high accuracy. Furthermore, \TheName{}-P can accelerate the training process (up to 61.2\% faster) without much accuracy degradation, which may even improve the accuracy because of the adapted or personalized local model.

\end{document}